\documentclass[a4paper,12pt]{article}

\usepackage{psfrag}
\usepackage{mathrsfs}

\usepackage{graphicx}
\usepackage[margin=15pt,font=small,labelfont=bf,labelsep=period]{caption}
\usepackage{amsmath}
\usepackage{amssymb}
\usepackage{amsfonts}
\usepackage{mathtools}
\usepackage{bm}
\usepackage{dsfont}
\usepackage{cite}
\usepackage{setspace}
\usepackage{environ,letltxmacro}
\usepackage{tikz}
\usetikzlibrary{arrows,decorations.pathreplacing}
\usepackage{enumitem}
\usepackage{hyperref}
\usepackage{relsize}
\usepackage[a4paper,textwidth=16.04cm,textheight=22cm,footskip=15mm]{geometry}

\numberwithin{equation}{section}

\LetLtxMacro\oldequation\equation
\LetLtxMacro\endoldequation\endequation
\let\equation\relax
\let\endequation\relax
\NewEnviron{equation}[1][\empty]{%
 \ifx \empty#1
  \oldequation \BODY \endoldequation
 \else
  \ifx b#1
   \oldequation \fbox{$\displaystyle \BODY $} \endoldequation
  \else
   \oldequation \BODY \endoldequation
  \fi
 \fi}
\newcommand{\e}{\mathrm{e}}
\newcommand{\Tr}{\operatorname{Tr}}
\newcommand{\STr}{\operatorname{STr}}
\newcommand{\Trp}{\operatorname{Tr^\prime}}
\newcommand{\detp}{\operatorname{det^\prime}}

\newcommand{\Rk}{\mathcal{R}_k}

\newcommand{\p}{\partial}
\newcommand{\sg}{\sqrt{g}}
\newcommand{\sgb}{\sqrt{\bar{g}}}
\newcommand{\sgbx}{\sqrt{\bar{g}(x)}}
\newcommand{\gb}{\bar{g}}
\newcommand{\hg}{\hat{g}}
\newcommand{\shg}{\sqrt{\hg}}
\newcommand{\shgx}{\sqrt{\hg(x)}}
\newcommand{\mn}{{\mu\nu}}
\newcommand{\rs}{{\rho\sigma}}
\newcommand{\mO}{\mathcal{O}}
\newcommand{\mD}{\mathcal{D}}
\newcommand{\mF}{\mathcal{F}}
\newcommand{\mA}{\mathcal{A}}
\newcommand{\mS}{\mathscr{S}}
\newcommand{\td}{\text{d}}
\newcommand{\dd}{\td^d}
\newcommand{\dex}{\td^{2+\varepsilon}x}
\newcommand{\hR}{\hat{R}}
\newcommand{\hD}{\hat{D}}
\newcommand{\hB}{\hat{\Box}}
\newcommand{\Ye}{Y_\varepsilon}

\newcommand{\vp}{\varphi}
\newcommand{\cg}{\check{g}}
\newcommand{\cl}{\check{\lambda}}

\newcommand{\ns}{N}
\newcommand{\UV}{\Lambda}

\newcommand{\SB}{S_\UV}

\newcommand{\bP}{\bar{\Phi}}
\newcommand{\hP}{\hat{\Phi}}
\newcommand{\hvp}{\hat{\varphi}}
\newcommand{\xb}{\bar{\xi}}
\newcommand{\Gg}{\Gamma_k^\text{grav}}
\newcommand{\GM}{\Gamma_k^\text{M}}
\newcommand{\gsc}{\bar{g}_k^\text{sc}}
\newcommand{\Asc}{\bar{A}_k^\text{sc}}
\newcommand{\ve}{\varepsilon}
\newcommand{\bG}{\breve{G}}
\newcommand{\bL}{\breve{\Lambda}}
\newcommand{\bGk}{\breve{G}_k}
\newcommand{\bgk}{\breve{g}_k}
\newcommand{\bLk}{\breve{\Lambda}_k}
\newcommand{\blk}{\breve{\lambda}_k}
\newcommand{\bgs}{\breve{g}_*}
\newcommand{\bls}{\breve{\lambda}_*}
\newcommand{\Ggd}{\Gamma_k^\text{grav,2D}}
\newcommand{\GgdN}{\Gamma_k^\text{grav,2D,NGFP}}
\newcommand{\GL}{\Gamma_k^\text{L}}
\newcommand{\tg}{\tilde{g}}
\newcommand{\stg}{\sqrt{\tg}}
\newcommand{\tR}{\tilde{R}}

\newcommand{\tB}{\tilde{\Box}}
\newcommand{\cgr}{c_\text{grav}^\text{NGFP}}
\newcommand{\SEH}{S_\text{EH}}
\newcommand{\mku}{\mkern1mu}
\newcommand{\NEH}{N_\text{EH}}
\newcommand{\gfull}{\e^{2\phi}\mku\hg}
\newcommand{\Zm}{Z_\text{matter}}
\newcommand{\Yg}{Y_\text{grav}^\text{NGFP}}
\newcommand{\Gi}{\Gamma_\text{ind}}
\newcommand{\ob}{\mathscr{O}}
\newcommand{\bob}{\bar{\mathscr{O}}}
\newcommand{\ZMc}{Z_\text{M}^{(c)}}
\newcommand{\mP}{\mathcal{P}}

\begin{document}

\begin{titlepage}

\title{
\begin{flushright}
\normalsize{MITP/15-113}
\bigskip
\vspace{1cm}
\end{flushright}
The unitary conformal field theory\\ behind 2D Asymptotic Safety\\[2mm]
}

\date{}

\author{Andreas Nink and Martin Reuter\\[3mm]
{\small Institute of Physics, PRISMA \& MITP,}\\[-0.2em]
{\small Johannes Gutenberg University Mainz,}\\[-0.2em]
{\small Staudingerweg 7, D--55099 Mainz, Germany}
}

\maketitle
\thispagestyle{empty}

\vspace{2mm}
\begin{abstract}
Being interested in the compatibility of Asymptotic Safety with Hilbert space positivity (unitarity), we consider a
local truncation of the functional RG flow which describes quantum gravity in $d>2$ dimensions and construct its limit
of exactly two dimensions. We find that in this limit the flow displays a nontrivial fixed point whose effective
average action is a non-local functional of the metric. Its pure gravity sector is shown to correspond to a unitary
conformal field theory with positive central charge $c=25$. Representing the fixed point CFT by a Liouville theory
in the conformal gauge, we investigate its general properties and their implications for the Asymptotic Safety program.
In particular, we discuss its field parametrization dependence and argue that there might exist more than one
universality class of metric gravity theories in two dimensions. Furthermore, studying the gravitational dressing in
2D asymptotically safe gravity coupled to conformal matter we uncover a mechanism which leads to a complete quenching
of the a priori expected Knizhnik--Polyakov--Zamolodchikov (KPZ) scaling. A possible connection
of this prediction to Monte Carlo results obtained in the discrete approach to 2D quantum gravity based upon causal
dynamical triangulations is mentioned. Similarities of the fixed point theory to, and differences from, non-critical
string theory are also described. On the technical side, we provide a detailed analysis of an intriguing connection
between the Einstein--Hilbert action in $d>2$ dimensions and Polyakov's induced gravity action in two dimensions.
\end{abstract}

\end{titlepage}

\newpage

\begin{spacing}{1.1}

\section{Introduction}
\label{sec:Intro}

During the past two decades, Asymptotic Safety \cite{W80} matured from a hypothetical scenario to a theory with a
realistic chance to describe the structure of spacetime and the gravitational interaction consistently and predictively
even on the shortest length scales possible. In particular, there is mounting evidence supporting the existence of the
decisive nontrivial renormalization group (RG) fixed point. Yet, a number of immediate questions are still open. The
most obvious one is about the precise nature of the action functional which describes this fixed point. In which way
exactly does it depend on the metric, the background metric, and the Faddeev--Popov ghosts? Is it local? What are the
structural properties of the fixed point theory, i.e.\ the one defined directly at the fixed point rather than by a
trajectory running away from it. Is this theory a conformal field theory?

In $2$ dimensions we are indeed used to the picture that the conformal field theories correspond to points in theory
space that are fixed points of the RG flow \cite{Na15}. In $4$ dimensions, however, Quantum Einstein Gravity (QEG) has
a scale invariant fixed point theory but it is unclear whether it is conformal.

While conformality is not known to be indispensable, there exist several other properties an asymptotically safe theory
\emph{must} possess in addition to its mere nonperturbative renormalizability, that is, the existence of a suitable
non-Gaussian fixed point. The two most important ones are clearly \emph{Background Independence} and \emph{unitarity}.
While there are by now first promising results which indicate that the requirements of Background Independence and
Asymptotic Safety can be met simultaneously in sufficiently general truncations of the RG flow \cite{BR14}, little
is known about the status of unitarity.

In this connection the somewhat colloquial term ``unitarity'' is equivalent to ``Hil\-bert space positivity'' and is
meant to express that the state space of the system under consideration contains no vectors having a negative scalar
product with itself (``negative norm states''). If it does so, it is not a Hilbert space in the mathematical sense of
the word and cannot describe a \emph{quantum} system as the probability interpretation of
quantum mechanics would break down then.

At least on (nondynamical) flat spacetimes the criterion of Hilbert space positivity, alongside with the spectral
condition can be translated from the Lorentzian to the Euclidean setting where it reappears as the requirement of
reflection-, or Osterwalder--Schrader, positivity \cite{OS73,S93,GJ87}.

Unitarity is indeed a property that is not automatic and needs to be checked in order to demonstrate the viability of
the Asymptotic Safety program based upon the Effective Average Action (EAA). The EAA for gravity \cite{R98} is a scale
dependent effective action for a BRST gauge-fixed theory. Its operator formulation amounts to an indefinite metric
(Krein space) quantization, and so the negative norm states it contains should be ``factored out'' ultimately in order
to obtain a positive (``physical'') state space, a true Hilbert space. While this procedure is standard and familiar
from perturbative quantum gravity and Yang--Mills theory, for instance, the situation is much more involved in
Asymptotic Safety. The reason is that, implicitly, this indefinite metric quantization is applied to a bare action
which is essentially given by the fixed point functional \cite{MR09,VZ11,M94,MS15}. As such it is already in itself the
result of a technically hard nonperturbative computation which in practice can be done only approximately, for the time
being.

In the present paper we explore the question of Hilbert space positivity together with a number of related issues such
as locality by analyzing the situation in $2$ dimensions where a number of technical simplifications occur. We start
out from the Einstein--Hilbert truncation of the EAA in $d=2+\ve>2$ dimensions, investigate the nature of its
$\ve\rightarrow 0$ limit, and finally construct a \emph{manifestly $2$-dimensional} action which describes 2D
Asymptotic Safety without reverting to ``higher'' dimensions in any way. In this manner we shall see that the
non-Gaussian fixed point (NGFP) underlying Asymptotic Safety is governed by a conformal field theory (CFT) which is
interesting in its own right, and whose properties we shall discuss. Interestingly enough, it turns out to possess a
positive central charge, thus giving rise to a unitary representation of the Virasoro algebra and a ``positive''
Hilbert space in the above sense.

\medskip
\noindent
\textbf{(1)}
A framework which allows to systematically search for asymptotically safe theories is the Effective Average Action, and
in particular its functional RG equation (FRGE): The RG flow of the EAA, $\Gamma_k$, is governed by the exact
nonperturbative evolution equation
\cite{RW93,W93,RW94,R98}
\begin{equation}
 k\p_k\Gamma_k = \frac{1}{2}\STr\left[\big(\Gamma_k^{(2)}+\Rk\big)^{-1}\,k\p_k\Rk\right] .
\label{eq:FRGE}
\end{equation}
It describes the dependence of $\Gamma_k$ on the RG scale $k$ which plays the role of a
regularization scale for infrared (IR) fluctuations. The suppression of IR modes is realized by the cutoff operator
$\Rk$, satisfying $\Rk\rightarrow k^2$ for IR and $\Rk\rightarrow 0$ for UV modes, respectively. Besides,
$\Gamma_k^{(2)}$ denotes the Hessian of the EAA with respect to the dynamical field. In
particular, in the case of gravity when the background field method is employed, it is the second functional derivative
of $\Gamma_k[g,\gb]$ w.r.t.\ $g$ at fixed background $\gb$. By construction, $\Gamma_k$ approaches the full quantum
effective action $\Gamma$ in the limit $k\rightarrow 0$. Although its $k\rightarrow\infty$ limit is \emph{formally}
related to the bare action \cite{W93,RW94}, in the Asymptotic Safety program no bare action is posited a priori; it is
rather ``reconstructed'' from the condition that this limit actually exists nonperturbatively
\cite{MR09,VZ11,M94,MS15}.

In order to find solutions to the FRGE \eqref{eq:FRGE} one usually resorts to truncations, implying a reduction of the
infinite dimensional theory space that consists of all possible action monomials compatible with the underlying
symmetry. By now a large variety of truncations have been studied to support the existence of a NGFP of metric gravity,
including for instance terms of higher orders in the curvature or actions that couple gravity to matter
\cite{NR06,P09,CPR09,RS12}.

\medskip
\noindent
\textbf{(2)}
In this paper we will focus on a gravity+matter system where the purely gravitational sector consists of the Euclidean
Einstein--Hilbert truncation,
\begin{equation}
 \Gamma_k^\text{grav}[g] = \frac{1}{16\pi G_k} \int \dd x \sg \,\big( -R + 2\mku\Lambda_k \big),
\label{eq:EHfunctional}
\end{equation}
and the matter contribution is given by a multiplet of $\ns$ scalar fields $A=(A^i)$, with $i=1,\cdots,\ns$, minimally
coupled to the full, dynamical metric:
\begin{equation}
 \GM[g,A]\equiv \Gamma^\text{M}[g,A]=\frac{1}{2}\sum\limits_{i=1}^{\ns}\int\dd x\sg\, g^\mn\,\p_\mu A^i\,\p_\nu A^i\,.
\label{eq:matter}
\end{equation}
Note that the matter action contains no running parameters in the present truncation(s).
Supplemented by appropriate gauge fixing and ghost terms, the sum of $\Gamma_k^\text{grav}$ and $\GM$ can be inserted
into the flow equation \eqref{eq:FRGE} in order to determine the running of Newton's constant $G_k$ and the
cosmological constant $\Lambda_k$.

The RG studies of this system for $d=2+\ve$ dimensions ($\ve\searrow 0$) are of particular importance for our
analysis. It is possible then to argue on general grounds that the $\beta$-function of the dimensionless Newton
constant $g_k\equiv G_k k^{d-2}=G_k k^\ve$ must be of the form
\begin{equation}
 \beta_g = \ve g - b g^2 + \mO(g^3)\,.
\end{equation}
This implies the existence of a nontrivial fixed point which is located at
\begin{equation}
 g_* = \ve/b\,,
\label{eq:NGFPWithB}
\end{equation}
up to higher $\ve$-orders. Therefore, we have $g_k\propto\ve$ and, in turn, $G_k\propto\ve$ in the vicinity of the
NGFP, having far-reaching consequences for the dimensional continuation of the Einstein--Hilbert action to two
dimensions.

\medskip
\noindent
\textbf{(3)}
In \emph{exactly} $2$ dimensions the integral $\int\td^2 x\sg\,R$ is a purely topological term according to the
Gauss--Bonnet theorem. In particular, it is independent of the metric and does not imply any local dynamics. Thus, one
might expect that \eqref{eq:EHfunctional} becomes trivial up to the cosmological constant term when $d$ approaches $2$.
However, the fact that the prefactor $1/G_k$ entails a $1/\ve$ pole gives so much weight to $\int\td^{2+\ve} x\sg\,R$
that the limit $\ve\rightarrow 0$ remains indeed nontrivial.

We will present a new argument in this work showing that
the (local) Einstein--Hilbert action turns into a \emph{non-local} action in the limit $d\rightarrow 2$. The essential
part of this limit action will be seen to be given by Polyakov's induced gravity action.

Our proof will confirm recurring speculation \cite{CD15} that the induced gravity action is the natural
$2$-dimensional analogue of the Einstein--Hilbert action in $d>2$ as both actions determine field equations for the
metric in their respective spacetime dimension.

\medskip
\noindent
\textbf{(4)}
Here we go one step further: We do not require that one action has to be replaced
by the other one when switching between $d=2$ and $d>2$. The idea is rather to say that there is \emph{only one} common
origin, the Einstein--Hilbert action in a general dimension $d$, and that \emph{the induced gravity action emerges
automatically when $d$ approaches $2$}.

It is this latter 2D action, analyzed at the NGFP, that establishes the contact between the Asymptotic Safety studies
within the Einstein--Hilbert truncation and $2$-dimensional conformal field theory. It will form the basis of our
investigations concerning central charges and unitarity.

\medskip
\noindent
\textbf{(5)}
It turns out that the crucial fixed point value $g_*$ depends on the way the metric is parametrized.
The background field technique which underlies the Asymptotic Safety studies expresses the dynamical metric $g_\mn$ in
terms of a fixed background metric $\gb_\mn$ and the fluctuations $h_\mn$. There are, however, different possible
choices of how these fields are related, in particular the standard linear parametrization $g_\mn=\gb_\mn+h_\mn$ and
the exponential parametrization $g_\mn=\gb_{\mu\rho}(\e^h)^\rho{}_\nu$. These two parametrizations
give rise to different $\beta$-functions and different central charges.
As we will argue, this disagreement does not necessarily mean that one choice is correct while the other one is wrong,
and we advocate the possibility that the two parametrizations might refer to different universality classes.
In the 2D limit, however, we provide strong evidence that the exponential parametrization is more appropriate for a
consistent description of the gravitational interactions. It is this choice that leads to a NGFP theory with the value
$c_\text{grav}=25$ for the central charge of the pure gravity sector.

\medskip
\noindent
\textbf{(6)}
This paper is organized as follows. We review Background Independence and the special role of self-consistent
backgrounds in \emph{section \ref{sec:BackgroundIndependence}}. In particular, we re-interpret the effective Einstein
equation as a tadpole condition and the trace of the stress energy tensor due to metric fluctuations as a kind of
classical ``trace anomaly''. Here, all calculations are performed in $2+\ve$ dimensions, and the 2D limit is taken
at the very end only.
This leads us to the question if the same trace anomaly can be obtained when starting out from a strictly 2D action.
The answer to this question will be given in \emph{section \ref{sec:IndGravityFromEH}} where we compute the 2D limit of
the Einstein--Hilbert action at the NGFP and argue that it results indeed in an action with the sought-for properties.
We demonstrate in \emph{section \ref{sec:NGFPCFT}} that this 2D gravitational NGFP action gives rise to a unitary
conformal field theory. Particular attention is paid to the relation between the crucial sign of its central charge,
the occurrence of a conformal factor instability, and unitarity. \emph{Section \ref{sec:Emergence}} is dedicated to
the impact of different metric parametrizations and establishes the special status of the exponential parametrization
in the 2D limit. Finally, \emph{section \ref{sec:recFI}} is devoted to the properties of the fixed point CFT, and
a comparison of Asymptotic Safety to other approaches to 2D gravity. For that purpose we construct a functional
integral which reproduces the effective average action related to the CFT behind the NGFP. This ``reconstructed''
functional integral is used to investigate the gravitational dressing of matter field operators when conformal matter
is coupled to asymptotically safe gravity. We demonstrate that, contrary to what one would have expected, there is
no KPZ scaling in our setting as a consequence of Asymptotic Safety. We also discuss similarities and differences
compared with non-critical string theory and Monte--Carlo simulations in the CDT approach.
\emph{Section \ref{sec:Conc}} contains our conclusions and an outlook.

In the appendix we catalog various useful identities for Weyl transformations, and we include a detailed discussion of
the induced gravity action in the presence of zero modes.

\section{Background Independence via background fields}
\label{sec:BackgroundIndependence}

In this preparatory section we collect a number of results concerning the implementation of Background Independence in
the EAA framework which actually does employ (unspecified!) background fields. In particular, we introduce the energy
momentum tensor of metric fluctuations in a background, as well as an associated ``trace anomaly''. The latter will be
used later on in order to identify the conformal field theory at the heart of Asymptotic Safety in $2$ dimensions.

\subsection{The effective Einstein equation re-interpreted}
\label{sec:EEE}

Let us consider a generic effective average action $\Gamma_k[\Phi,\bP]\equiv \Gamma_k[\vp;\bP]$ involving a multiplet of
dynamical fields $\big\langle\hP^i\big\rangle\equiv \Phi^i$, associated background fields $\bP^i$, and fluctuations $\vp^i
\equiv \langle \hvp^i\rangle = \Phi^i - \bP^i$. The effective average action implies a source $\leftrightarrow$ field
relationship which contains an explicit cutoff term linear in the fluctuation fields:
\begin{equation}
 \frac{1}{\sgb}\frac{\delta\Gamma_k[\vp;\bP]}{\delta\vp^i(x)}+\Rk[\bP]_{ij}\,\vp^j(x)=J_i(x) \,.
\label{eq:so-fi}
\end{equation}
By definition, self-consistent backgrounds are field configurations $\bP(x)\equiv \bP_k^\text{sc}(x)$ which allow
$\vp^i=0$ to be a solution of \eqref{eq:so-fi} with $J_i=0$. A self-consistent background is particularly ``liked'' by
the fluctuations, in the sense that they leave it unaltered on average: $\langle\hP\rangle = \bP+\langle\hvp\rangle =
\bP^\text{sc}$. These special backgrounds are determined by the tadpole condition $\langle\hvp^i\rangle=0$, which reads
explicitly
\begin{equation}
 \frac{\delta}{\delta\vp^i(x)}\Gamma_k[\vp;\bP]\Big|_{\vp=0,\,\bP=\bP_k^\text{sc}} = 0 \,.
\label{eq:tapo}
\end{equation}
Equivalently, in terms of the full dynamical field,
\begin{equation}
 \frac{\delta}{\delta\Phi^i(x)}\Gamma_k[\Phi,\bP]\Big|_{\Phi=\bP=\bP_k^\text{sc}} = 0 \,.
\label{eq:tapo-full}
\end{equation}

In this paper we consider actions of the special type
\begin{equation}
 \Gamma_k[g,\xi,\xb,A,\gb] = \Gg[g,\gb] +\GM[g,A,\gb] +\Gamma_k^\text{gf}[g,\gb] +\Gamma_k^\text{gh}[g,\xi,\xb,\gb] .
\end{equation}
These functionals include a purely gravitational piece, $\Gg$, furthermore a (for the time being) generic matter action
$\GM$, as well as gauge fixing and ghost terms, $\Gamma_k^\text{gf}$ and $\Gamma_k^\text{gh}$, respectively. Concerning the
latter, only the following two properties are needed at this point:
(i) The $h_\mn$-derivative of the gauge fixing functional $\Gamma_k^\text{gf}[h;\gb] \equiv 
\Gamma_k^\text{gf}[\gb+h,\gb]$ vanishes at $h_\mn=0$. This is the case, for example, for classical gauge fixing terms
$S_\text{gf} \propto \int(\mF h)^2$ which are quadratic in $h_\mn$.
(ii) The functional $\Gamma_k^\text{gh}$ is ghost number conserving, i.e.\ all terms contributing to it have an equal
number of ghosts $\xi^\mu$ and antighosts $\xb_\mu$. Again, classical ghost kinetic terms
$\propto \int\xb\mathcal{M}\xi$ are of this sort.

Thanks to these properties, $\Gamma_k^\text{gf}$ drops out of the tadpole equation \eqref{eq:tapo-full}, and it
follows that $\xi=0=\xb$ is always a consistent background for the Faddeev--Popov ghosts. Adopting this background
for the ghosts, \eqref{eq:tapo-full} boils down to the following conditions for self-consistent metric and matter
field configurations $\gsc$ and $\Asc$, respectively:
\begin{align}
 0 &= \frac{\delta}{\delta g_\mn(x)}\Big\{\Gg[g,\gb]+\GM[g,\Asc,\gb]\Big\}\Big|_{g=\gb=\gsc} \;,
 \label{eq:selfcon} \\
 0 &= \frac{\delta}{\delta A(x)}\GM[g,A,\gb]\Big|_{g=\gb=\gsc,\, A=\Asc} \;.
 \label{eq:selfconmatter}
\end{align}
Introducing the energy-momentum tensor of the matter field,
\begin{equation}
 T^\text{M}[\gb,A]^\mn(x) \equiv \frac{2}{\sgbx} \frac{\delta}{\delta g_\mn(x)} \GM[g,A,\gb]\Big|_{g=\gb} \;,
\end{equation}
the first condition, equation \eqref{eq:selfcon}, becomes
\begin{equation}
 0 = \frac{2}{\sgbx} \frac{\delta}{\delta g_\mn(x)} \Gg[g,\gb]\Big|_{g=\gb=\gsc} + T^\text{M}[\gsc,\Asc]^\mn(x) .
\label{eq:EFE}
\end{equation}
This relation plays the role of an effective gravitational field equation which, together with the matter
equation \eqref{eq:selfconmatter}, determines $\gsc$ and $\Asc$. Structurally, eq.\ \eqref{eq:EFE} is a generalization
of the classical Einstein equation to which it reduces if $\Gg[g,\gb]\equiv \Gg[g]$ happens to have no ``extra
$\gb$-dependence'' \cite{MR10} and to coincide with the Einstein--Hilbert action; then the
$\delta/\delta g_\mn$-term in \eqref{eq:EFE} is essentially the Einstein tensor $G_\mn$.

In this very special background-free case we recover the familiar setting of classical General Relativity where there is
a clear logical distinction between matter fields and the metric, meaning the full one, $g_\mn$, while none other appears
in the fundamental equations then. It is customary to express this distinction by putting $G_\mn$ on the LHS of Einstein's
equation, the side of gravity, and $T_\mn^\text{M}$ on the RHS, the side of matter.

In the effective average action approach where, for both deep conceptual and technical reasons \cite{MR10,BR14}, the
introduction of a background is unavoidable during the intermediate calculational steps, this categorical distinction of
matter and gravity, more precisely, matter fields and metric fluctuations, appears unmotivated. It is much more natural to
think of $h_\mn$ as a \emph{matter} field which propagates on a background spacetime furnished with the metric $\gb_\mn$.

Adopting this point of view, we interpret the $\delta/\delta g_\mn$-term in \eqref{eq:EFE} as the energy-momentum tensor of
the $h_\mn$-field, and we define
\begin{equation}[b]
 T^\text{grav}[\gb]^\mn(x) \equiv \frac{2}{\sgbx} \frac{\delta}{\delta g_\mn(x)} \Gg[g,\gb]\Big|_{g=\gb}
 = \frac{2}{\sgbx} \frac{\delta}{\delta h_\mn(x)} \Gg[h;\gb]\Big|_{h=0} \;.
\end{equation}
The tadpole equation \eqref{eq:EFE} turns into an Einstein equation with zero LHS then:
\begin{equation}[b]
 0 = T_\mn^\text{grav}[\gsc] + T_\mn^\text{M}[\gsc,\Asc] .
\label{eq:Tadpole}
\end{equation}
It says that for a background to be self-consistent, the total energy-momentum tensor of matter and metric
fluctuations, in this background, must vanish. (In the general case there could be a contribution from the
ghosts also.)

\subsection[The stress tensor of the \texorpdfstring{$h_\mn$}{h}-fluctuations]
 {The stress tensor of the \texorpdfstring{$\bm{h_\mn}$}{h}-fluctuations}
\label{sec:Stresshmunu}

Note that in general, $T_\mn^\text{grav}$ is not conserved, $\bar{D}_\mu T^\text{grav}[\gb]^\mn\neq 0$, since due to the
presence of two fields in $\Gg$ the standard argument does noes not apply. Of course, it is conserved in the special
case $\Gg[g,\gb]\equiv\Gg[g]$ when there is no extra $\gb$-dependence.

For example, choosing $\Gg[\gb]$ to be the single-metric Einstein--Hilbert functional \eqref{eq:EHfunctional},
the corresponding energy-momentum tensor of the $h_\mn$-fluctuations is given by the divergence-free expression
\begin{equation}
 T_\mn^\text{grav}[\gb] = \frac{1}{8\pi G_k} \Big(\bar{G}_\mn + \Lambda_k\, \gb_\mn \Big) ,
\label{eq:EHEMTensor}
\end{equation}
with $\bar{G}_\mn$ the Einstein tensor built from $\gb_\mn$. The trace of the energy-momentum tensor
\eqref{eq:EHEMTensor} reads
\begin{equation}
 \Theta_k[\gb] \equiv \gb^\mn\, T_\mn^\text{grav}[\gb] = \frac{1}{16\pi G_k}\Big[ -(d-2)\bar{R}+2d\,\Lambda_k \Big],
\end{equation}
where $\bar{R}\equiv R(\gb)$. A remarkable feature of this trace is that it possesses a completely well defined,
unambiguous limit $d\rightarrow 2$ if $G_k$ and $\Lambda_k$ are of first order in $\ve=d-2$. In terms of the finite
quantities $\bGk \equiv G_k/\ve$ and $\bLk \equiv \Lambda_k/\ve$ which are of order $\ve^0$, we have
\begin{equation}
 \begin{split}
  \Theta_k[\gb] &= \frac{1}{16\pi\bGk} \Big[-\bar{R}+4\bLk\Big] +\mO(\ve) \\
    &= \frac{1}{16\pi\bgk} \Big[-\bar{R}+4k^2\blk\Big] +\mO(\ve).
 \end{split}
\label{eq:Theta2}
\end{equation}
In the second line of \eqref{eq:Theta2} we exploited that in exactly two dimensions the dimensionful and
dimensionless Newton constant are equal, so $g_k = G_k$ and $\bgk = \bGk$, while, as always,
$\lambda_k \equiv \Lambda_k/k^2$, hence $\blk=\bLk/k^2$.

When the underlying RG trajectory is in the NGFP scaling regime the dimensionless couplings are scale
independent, and
\begin{equation}
 \Theta_k^\text{NGFP}[\gb] = \frac{1}{16\pi\breve{g}_*} \Big[-\bar{R}+4\breve{\lambda}_* k^2 \Big].
\end{equation}
Using the parametrization $g_*\equiv \ve/b$ as in references \cite{NR13,N15,BR14} we obtain
\begin{equation}[b]
 \Theta_k^\text{NGFP}[\gb] = \Big({\textstyle\frac{3}{2}}b\Big) \, \frac{1}{24\pi}
  \Big[-\bar{R}+4\breve{\lambda}_* k^2 \Big].
\label{eq:Theta3}
\end{equation}
Here and in the following we consider $\Theta_k$ and $\Theta_k^\text{NGFP}$ as referring to \emph{exactly $2$
dimensions}, in the sense that the limit has already been taken, and we omit the ``$\mO(\ve)$'' symbol.

\subsection{The intrinsic description in exactly 2 dimensions}
\label{sec:descTwoD}

In this paper we would like to describe the limit $d\rightarrow 2$ of QEG in an intrinsically $2$-dimensional
fashion, that is, in terms of a new functional $\Gamma_k^\text{grav,2D}$ whose arguments are fields in strictly
$2$ dimensions, and which no longer makes reference to its ``higher'' dimensional origin. Since the Einstein--Hilbert
term is purely topological in exactly $d=2$, it is clear that the sought-for action must have a different structure.

\medskip
\noindent
\textbf{(1)}
One of the conditions which we impose on $\Gamma_k^\text{grav,2D}$ is that it must reproduce the trace $\Theta_k$
computed in $d>2$, since we saw that this quantity has a smooth limit with an immediate interpretation in $d=2$ exactly:
\begin{equation}
 2 g_\mn \frac{\delta}{\delta g_\mn} \Gamma_k^\text{grav,2D}[g,\gb]\Big|_{g=\gb}=\sgb\, \Theta_k[\gb].
\label{eq:RelEMTensors}
\end{equation}
Furthermore, if $\Gg$ is a single-metric action, we assume that $\Gamma_k^\text{grav,2D} \equiv \Gamma_k^\text{grav,2D}[g]$
has no extra $\gb$-dependence either. The condition \eqref{eq:RelEMTensors} fixes its response to an infinitesimal Weyl
transformation then:
\begin{equation}
 2 g_\mn(x) \frac{\delta}{\delta g_\mn(x)} \Gamma_k^\text{grav,2D}[g] 
 \equiv \frac{\delta}{\delta \sigma(x)} \Gamma_k^\text{grav,2D}[\e^{2\sigma} g]\Big|_{\sigma=0}
 = \sqrt{g(x)}\,\Theta_k[g](x).
\label{eq:InfWeyl}
\end{equation}
For the example of the Einstein--Hilbert truncation, $\Theta_k$ is of the form
\begin{equation}
 \Theta_k[g]=a_1(-R+a_2),
\label{eq:ThetaEH} 
\end{equation}
with constants $a_1,a_2$ which can be read off from \eqref{eq:Theta2} -- \eqref{eq:Theta3} for the various cases.

\medskip
\noindent
\textbf{(2)}
It is well known how to integrate equation \eqref{eq:InfWeyl} in the conformal gauge \cite{P81}. By setting
\begin{equation}
 g_\mn(x) = \hg_\mn(x)\,\e^{2\phi(x)},
\end{equation}
with a fixed reference metric $\hg_\mn$ (conceptually unrelated to $\gb_\mn$), one for each topological sector, and
taking advantage of the identities in appendix \ref{app:Weyl}, eq.\ \eqref{eq:InfWeyl} with \eqref{eq:ThetaEH} is
turned into
\begin{equation}
 \frac{\delta}{\delta \phi(x)} \Ggd[\e^{2\phi} \hg] = a_1\shgx \Big[2\hD_\mu\hD^\mu\phi(x)-\hR(x)
  +a_2 \,\e^{2\phi(x)}\Big].
\end{equation}
The general solution to this equation is easy to find:
\begin{equation}[b]
 \Ggd[\hg\e^{2\phi}] = \GL[\phi;\hg]+U_k[\hg].
\label{eq:GravToLiou}
\end{equation}
Here $U_k$ is a completely arbitrary functional of $\hg$, independent of $\phi$, and $\GL$ denotes the Liouville action
\cite{RW97}:
\begin{equation}
 \begin{split}
	\GL[\phi;\hg] &= (-2a_1)\int\td^2x\shg \left({\frac{1}{2}}\hD_\mu\phi\hD^\mu\phi
	  + {\frac{1}{2}}\hR\phi - {\frac{a_2}{4}}\e^{2\phi} \right) \\
		&= (-2a_1)\,\Delta I[\phi;\hg] + \frac{1}{2}a_1 a_2\int\td^2x\shg\,\e^{2\phi} \,.
 \end{split}
\label{eq:LiouvilleAction}
\end{equation}
In the last line we employed the normalized functional
\begin{equation}
 \Delta I[\phi;g] \equiv \frac{1}{2}\int\td^2x\sg\,\big(D_\mu\phi D^\mu\phi+R\phi\big).
\label{eq:DeltaIDef}
\end{equation}

While this method of integrating the trace ``anomaly'' applies in all topological sectors, it is \emph{unable to find
the functional} $U_k[\hg]$. Usually in conformal field theory or string theory this is not much of a disadvantage,
but in quantum gravity where Background Independence is a pivotal issue it is desirable to have a more complete
understanding of $\Ggd$. For this reason we discuss in the next section the possibility of taking the limit
$\ve\rightarrow 0$ directly at the level of the action.

\section{How the Induced Gravity Action emerges from the Einstein--Hilbert action}
\label{sec:IndGravityFromEH}

In this section we reveal a mechanism which allows us to regard Polyakov's induced gravity action in $2$ dimensions as
the $\ve\rightarrow 0$ limit of the Einstein--Hilbert action in $2+\ve$ dimensions. Here and in the following
we always consider the case $\ve>0$, i.e.\ the limit $\ve\searrow 0$. This will confirm the point of view that the
induced gravity action is fundamental in describing $2$-dimensional gravity, while it is less essential for
$d>2$ where gravity is governed mainly by an (effective average) action of Einstein--Hilbert type. The
dimensional limit exhibits a discontinuity at $d=2$, producing a non-local action out of a local one.

\medskip
\noindent
\textbf{(1)}
The crucial ingredient for a nontrivial limit $\ve\rightarrow 0$ is a prefactor of the Einstein--Hilbert action
proportional to $1/\ve$. This occurs whenever the Newton constant is proportional to $\ve$. As mentioned in the
introduction, such a behavior was found in the Asymptotic Safety related RG studies. The renormalization group flow has
a non-Gaussian fixed point with a Newton constant of order $\ve$; a result which is independent of the underlying
regularization scheme and which is found in both perturbative and nonperturbative investigations.

Employing a reconstruction formula \cite{MR09,MS15,NR16} we shall see that this property holds not only for the
effective, but also for the \emph{bare} action: Using an appropriate regularization prescription the bare Newton
constant is of first order in $\ve$, too.

This is our motivation for considering a generic Einstein--Hilbert action with a Newton constant proportional to $\ve$.
For the discussion in this section it is not necessary to specify the physical role of the action under
consideration -- the arguments apply to both bare and effective (average) actions. In both cases our aim is eventually
to make sense of, and to calculate
\begin{equation}
 \frac{1}{\ve}\int\td^{2+\ve}x\sg\,R
\label{eq:LimitInt}
\end{equation}
in the limit $\ve\rightarrow 0$.

\medskip
\noindent
\textbf{(2)}
It turns out helpful to consider the transformation behavior of the Einstein--Hilbert action under Weyl
rescalings. In this way an expansion in powers of $\ve$ is more straightforward. Loosely speaking, the reason why Weyl
variations are related to the 2D limit resides in the fact that the conformal factor is the only dynamical part of the
metric that ``survives'' when the limit $d\rightarrow 2$ is taken, i.e.\ the conformal sector captures the essential
information in $d=2+\ve$. This idea is detailed in subsection \ref{sec:ConfGauge}.

Weyl transformations are defined by the pointwise rescaling
\begin{equation}
 g_\mn(x) = \e^{2\sigma(x)} \hg_\mn(x) \,,
\label{eq:WeylTrans}
\end{equation}
with $\sigma$ a scalar function on the spacetime manifold. In appendix \ref{app:Weyl} we list the transformation
behavior of all quantities relevant in this section.

From \eqref{eq:WeylTrans} it follows that $g_\mn$ is invariant under the split-symmetry transformations
\begin{equation}
\hg_\mn\rightarrow \e^{2\chi}\hg_\mn \, ,\qquad \sigma\rightarrow \sigma - \chi \, .
\label{eq:SplitSymmetry}
\end{equation}
Thus, any functional of the full metric $g_\mn$ rewritten in terms of $\hg_\mn$ and $\sigma$ is invariant under
\eqref{eq:SplitSymmetry}. On the other hand, a functional of $\hg_\mn$ and $\sigma$ which is not split symmetry
invariant cannot be expressed as a functional involving only $g_\mn$, but it contains an ``extra $\hg_\mn$-dependence''
\cite{MR10}.

Before actually calculating the 2D limit of \eqref{eq:LimitInt} in section \ref{sec:EpsilonLimit} and
\ref{sec:LimitFullEH} in a gauge invariant manner, we illustrate the situation in section \ref{sec:ConfGauge} by
employing the conformal gauge, and we give some general arguments in section \ref{sec:genRem} why and in what sense
the limit is well defined.

\subsection{Lessons from the conformal gauge}
\label{sec:ConfGauge}

In exactly $2$ spacetime dimensions any metric $g$ can be parametrized by a diffeomorphism $f$ and a Weyl scaling
$\sigma$:
\begin{equation}
 f^* g = \e^{2\sigma}\,\hg_{\{\tau\}} \,,
\label{eq:MetricDiffWeyl}
\end{equation}
where $f^*g$ denotes the pullback of $g$ by $f$, and $\hg_{\{\tau\}}$ is a fixed reference metric that depends only on
the Teichm\"uller parameters $\{\tau\}$ or ``moduli'' \cite{IT92}. Stated differently, a combined Diff$\times$Weyl
transformation can bring any metric to a reference form. Thus, the moduli space is the remaining space of inequivalent
metrics, $\mathcal{M}_h= \mathcal{G}_h/ (\text{Diff}\times\text{Weyl})_h$, where $\mathcal{G}_h$ is the space of all
metrics on a genus-$h$ manifold.\footnote{For the topology of a sphere $\mathcal{M}_h=\mathcal{M}_0$ is trivial, while
for a torus there is one complex parameter, $\tau$, assuming values in the fundamental region, $F_0$. Apart from such
simple examples it is notoriously involved to find moduli spaces \cite{IT92}.} Its precise form is irrelevant for the
present discussion. Accordingly, if not needed we do not write down the dependence on $\{\tau\}$ explicitly in the
following. Here we consider $\hg$ a reference metric for a fixed topological sector.

In order to cope with the redundancies stemming from diffeomorphism invariance we can fix a gauge by picking one
representative among the possible choices for $f$ in eq.\ \eqref{eq:MetricDiffWeyl}, the most natural choice being
the conformal gauge:
\begin{equation}
 g_\mn = \e^{2\sigma}\,\hg_\mn \,.
\label{eq:ConfGauge}
\end{equation}
Equation \eqref{eq:ConfGauge} displays very clearly the special role of $2$ dimensions: The metric depends only on the
conformal factor and possibly on some topological moduli parameters. Since the latter are global parameters, we see
that \emph{locally} the metric is determined only by the conformal factor.

\medskip
\noindent
\textbf{(1) Conformal flatness}.
At this point a comment is in order. By choosing an appropriate coordinate system it is always possible to bring a
2D metric to the form
\begin{equation}
 g_\mn = \e^{2\sigma} \delta_\mn\,,
\label{eq:NotConfGauge}
\end{equation}
in the neighborhood of an arbitrary spacetime point, where $\delta_\mn$ is the flat Euclidean metric (see ref.\
\cite{DFN92} for instance). However, this is a local property only. \emph{For a general metric on a general 2D manifold
there exists no scalar function $\sigma$ satisfying \eqref{eq:NotConfGauge} globally}.\footnote{This can be understood
by means of the following counterexample. Consider the standard sphere $S^2\subset \mathbb{R}^3$ with the induced
metric. Upon stereographic projection the sphere is parametrized by isothermal coordinates, say $(u,v)$, where the
metric assumes the form $g=\frac{4}{(1+u^2+v^2)^2} (\td u^2+\td v^2)$. Setting $\sigma\equiv\ln\left(\frac{2}{1+
u^2+v^2}\right)$ we have $g=\e^{2\sigma}\hg$ with $\hg=\delta$. If we assumed that $g=\e^{2\sigma}\hg$ holds globally
for a valid scalar function $\sigma$, we could make use of identity \eqref{eq:WeylgR} to arrive at a contradiction for
the Euler characteristic $\chi=2\mku$, namely: $8\pi=4\pi\chi\equiv\int\!\sg\,R = \int\!\shg\,(\hR-2\,
\hB\mku\sigma)=-2\int\!\shg\;\hB\mku\sigma=0$, since $\hR=0$ for the flat metric, and since the sphere has vanishing
boundary. A resolution to this contradiction is to take into account that we need (at least) two coordinate patches all
of which have a boundary contributing to $\int\!\sg R$. Decomposing $S^2$ into two half spheres, $H_+$ and $H_-$, for
instance, and using $\hB\mku\sigma=-4/(1+u^2+v^2)^2$, we obtain $\int\!\sg\,R
=-2\int_{H_+}\shg\;\hB\mku\sigma-2\int_{H_-}\shg\;\hB\mku\sigma =8\pi=4\pi\chi$, as it should be.}
Rather must the reference metric in eq.\ \eqref{eq:ConfGauge} be compatible with all topological constraints, e.g.\ the
value of the integral $\int\shg\,\hR$ which is fixed by the Euler characteristic, a topological invariant that measures
the number of handles of the manifold. As a consequence, we cannot restrict our discussion to a globally conformally
flat metric in general.

\medskip
\noindent
\textbf{(2) \bm{$\text{Diff}\times\text{Weyl}$} invariant functionals}.
This has a direct impact on diffeomorphism and Weyl invariant functionals $F:g\mapsto F[g]$. The naive argument
claiming that diffeomorphism invariance can be exploited to make $g_\mn$ conformally flat, and then Weyl invariance to
bring it to the form $\delta_\mn$ such that $F[g]=F[\delta]$ would be independent of the metric, i.e.\ constant, is
wrong actually. The global properties of the manifold thwart this argument.

When choosing appropriate local coordinates
to render $g$ flat up to Weyl rescaling, there is some information of the metric implicitly encoded in the coordinate
system, e.g.\ in the boundary of each patch, giving rise to a remaining metric dependence in $F$. A combined
Diff$\times$Weyl transformation can bring the metric to unit form, but it changes boundary conditions (like periodicity
constraints for a torus) as well (see e.g.\ \cite{P98}). Therefore, $F$ is in fact constant with respect to local
properties of the metric, while it can still depend on global parameters. According to eq.\ \eqref{eq:MetricDiffWeyl}
these are precisely the moduli parameters. Hence, \emph{the metric dependence of any 2D functional which is both
diffeomorphism and Weyl invariant is reduced to a dependence on} $\{\tau\}$, and we can write $F[g]=f\big(\{\tau\}
\big)$ where $f$ is a function (not a functional).

\medskip
\noindent
\textbf{(3) Calculating 2D limits}.
Let us come back to the aim of this section, simplifying calculations by employing the conformal gauge
\eqref{eq:ConfGauge}. Following the previous discussion we should not rely on the choice \eqref{eq:NotConfGauge}.
Nevertheless, as an example we may assume for a moment that the manifold's topology is consistent with a metric $\hg$
that corresponds to a flat space, where -- for the above reasons -- conformal flatness is not expressed in local
coordinates as in \eqref{eq:NotConfGauge} but by the coordinate free condition $\hR=0$, which is possible iff the Euler
characteristic vanishes. The general case with arbitrary topologies will be covered in section \ref{sec:EpsilonLimit}.
We now aim at finding a scalar function $\sigma$ which is compatible with eq.\ \eqref{eq:ConfGauge} with $g_\mn$ given.
Exploiting the identities \eqref{eq:WeylLaplace} and \eqref{eq:WeylR} given in the appendix with $\hR=0$ we obtain
\begin{equation}
 R=-2\,\Box\mku\sigma \,.
\label{eq:CondSigma}
\end{equation}
Once we have found a solution $\sigma$ to eq.\ \eqref{eq:CondSigma}, it is clear that $\sigma'=\sigma+
\text{(\emph{zero modes of} }\Box\text{)}$ defines a solution, too. In particular, we can subtract from $\sigma$ its
projection onto the zero modes. This way, we can always obtain a solution to \eqref{eq:CondSigma} which is free of zero
modes. Thus, we can assume that $\sigma$ does not contain any zero modes before actually having computed it. In doing
so, relation \eqref{eq:CondSigma} can safely be inverted (cf.\ appendix \ref{app:Weyl} for a more detailed discussion
of zero modes):
\begin{equation}
 \sigma = -\frac{1}{2}\,\Box^{-1}R
\label{eq:SigmaSolved}
\end{equation}
Note that the possibility of performing such a direct inversion is due to the simple structure of eq.\
\eqref{eq:CondSigma} which, in turn, is a consequence of $\hR=0$.

Now we leave the strictly $2$-dimensional case and try to ``lift'' the discussion to $d=2+\ve$. For this purpose we
make the assumption that we can still parametrize the metric by \eqref{eq:ConfGauge} with a reference metric $\hg$
whose associated scalar curvature vanishes, $\hR=0$. (Once again, the general case will be discussed in section
\ref{sec:EpsilonLimit}.) In this case, by employing equations \eqref{eq:EHexpanded} and \eqref{eq:WeylLaplace} we
obtain the following relation for the integral \eqref{eq:LimitInt}:
\begin{equation}
 \frac{1}{\ve}\int\td^{2+\ve}x\sg\,R=\frac{1}{\ve}\int\td^2 x\shg\,\Big[\ve\,\sigma\big(-\hB\big)\sigma\Big]+\mO(\ve).
\end{equation}
This expression can be rewritten by means of the $2+\ve$-dimensional analogues of eqs.\ \eqref{eq:CondSigma} and
\eqref{eq:SigmaSolved} which read $R=-2\,\Box\mku\sigma+\mO(\ve)$ and $\sigma = -\frac{1}{2}\,\Box^{-1}R+\mO(\ve)$,
respectively, and we arrive at the result
\begin{equation}
\phantom{(\hR=0)}\quad
\fbox{$\displaystyle
 \frac{1}{\ve}\int\td^{2+\ve}x\sg\,R = -\frac{1}{4}\int\td^2 x\sg\, R\,\Box^{-1} R +\mO(\ve).
$}\quad (\hR=0)
\end{equation}

Clearly, the assumption $\hR=0$ is quite restrictive. But already in this simple setting we make a crucial
observation: the emergence of a non-local action from a purely local one in the limit $d\rightarrow 2$. More
precisely, \emph{in the 2D limit the Einstein--Hilbert type action} $\frac{1}{\ve}\int\td^{2+\ve}x\sg\,R$ \emph{becomes
proportional to the induced gravity action}. As we will see below, a similar result is obtained for general topologies
without any assumption on $\hR$.

\subsection{General properties of the limit}
\label{sec:genRem}

\textbf{(1) Existence of the limit}. In the following we argue that $\lim_{\ve\rightarrow 0}\left(\frac{1}{\ve}
\int\td^{2+\ve}x\,\sg\,R\right)$ is indeed a meaningful quantity without restricting ourselves to a particular
topology or gauge. For convenience let us set
\begin{equation}
 \mS_\ve[g] \equiv \int\td^{2+\ve}x\,\sg\, R .
\label{eq:SEpsilon}
\end{equation}
We would like to establish that $\mS_\ve[g]$ has a Taylor series in $\ve$ whose first nonzero term which is sensitive
to the \emph{local} properties of $g_\mn$ is of order $\ve$.

For the proof we make use of the relation $R_\mn=\frac{1}{2}g_\mn R$, valid in $d=2$ for any metric, so that the
Einstein tensor vanishes identically in $d=2$,
\begin{equation}
 G_\mn\big|_{d=2}=0 \, .
\end{equation}
Going slightly away from $2$ dimensions, $d=2+\ve$, we assume continuity and thus
conclude that $G_\mn\big|_{d=2+\ve}=\mO(\ve)$. Furthermore, the order $\ve^1$ is really the first
nonvanishing term of the Taylor series with respect to $\ve$ in general, i.e.\ $G_\mn\big|_{d=2+\ve}$ is not of
order $\mO(\ve^2)$ or higher. This can be seen by taking the trace of $G_\mn$,
\begin{equation}
 g^\mn G_\mn=g^\mn\left(R_\mn-\frac{1}{2}g_\mn R\right)=R-\frac{d}{2}R = -\frac{1}{2}R\,\ve.
\end{equation}
Therefore, we have $G_\mn \propto \ve$ (of course we assume $R\neq 0$ since $\mS_\ve$ would vanish identically
otherwise), and, similarly $G^\mn\propto\ve$. But even the non-trace (tensor) parts
of $G_\mn$ can be expected to be of order $\ve$ in general, as the following argument suggests. Let us consider a Weyl
transformation of the metric, $g_\mn=\e^{2\sigma}\hg_\mn$. The corresponding transformation of the Einstein tensor is given
by equation \eqref{eq:WeylEinstein} in the appendix. Now, let us assume that $\hg_\mn$ belongs to an Einstein manifold,
i.e.\ to a maximally symmetric spacetime.\footnote{In $d>2$ it is always possible to find $\sigma$ for a given metric $g_\mn$
such that $\hg_\mn=\e^{-2\sigma}g_\mn$ leads to a space with constant curvature provided that the manifold is compact. This
is known as the Yamabe problem \cite{Y60} (while the case $d=2$ is covered by Poincar\'{e}'s uniformization theorem).
However, this statement does not imply that the manifold is Einstein. In fact, there are known examples of metrics which are
not conformal to any Einstein metric \cite{NP01}. On the other hand, in $d=2$ any Riemannian manifold is of Einstein type.} In
this case the Ricci tensor is proportional to the scalar curvature, $\hR_\mn = \frac{1}{d}\hg_\mn\hR$. Then the Einstein
tensor reads
\begin{equation}
 G_\mn = (d-2)\left[-\frac{1}{2d}\hg_\mn\hR -\hD_\mu \hD_\nu \sigma + \hg_\mn\hB\sigma + \hD_\mu\sigma \hD_\nu\sigma
  + \frac{d-3}{2}\hg_\mn \hD_\alpha\sigma \hD^\alpha\sigma \right],
\end{equation}
so we find $G_\mn\propto\ve$ again.

This $\ve$-proportionality is exploited now to make a statement about the Taylor series of $\mS_\ve$. For that purpose we
consider the variation of $\mS_\ve$ with respect to $g_\mn$ (assuming vanishing surface terms):
\begin{equation}
 \begin{split}
  \frac{\delta \mS_\ve[g]}{\delta g_\mn(x)}= \int\td^{2+\ve}y\,\sg\left[\frac{1}{2}g^\mn R -R^\mn\right]\delta(x-y)
  = -\sg \, G^\mn = \mO(\ve).
 \end{split}
\label{eqn:SVariation}
\end{equation}
As a result we obtain $\mS_\ve[g]=C+\mO(\ve)$, where the constant $C$ is independent of $g_\mn$. Clearly, $C$
is obtained by computing $\mS_\ve$ in $d=2$, which is known to lead to the Euler characteristic $\chi\,$:
\begin{equation}
 C=\mS_\ve\big|_{\ve=0}=4\pi\chi.
\end{equation}
That is, we have $\mS_\ve=4\pi\chi+\mO(\ve)$. (This result differs from ref.\ \cite{CK80}, but it is in agreement with
refs.\ \cite{MR93,J06,GJ10}). As a consequence, the integral \eqref{eq:LimitInt} amounts to
\begin{equation}[b]
 \frac{1}{\ve}\int\td^{2+\ve}x\sg\, R = \frac{4\pi\chi}{\ve} + \text{finite}
 = \text{top.} + \text{finite},
\label{eq:expansion}
\end{equation}
where 'top.' is a field independent (up to topological information) and thus irrelevant contribution to the action.
The terms in \eqref{eq:expansion} that contain the interesting information about the dynamics of the field are of order
$\mO(\ve^0)$, so the ``relevant'' part of $\frac{1}{\ve}\int\td^{2+\ve}x\,\sg\, R$ has indeed a
meaningful limit $\ve\rightarrow 0$.

\medskip
\noindent
\textbf{(2) The role of the volume form}. Next we argue that the important part of the $\ve$-dependence of
$\mS_\ve$ originates from the scalar density $\sg\,R\,$ in the integrand of \eqref{eq:SEpsilon} alone, i.e.\ loosely
speaking it is sufficient to employ the a priori undefined fractional integration element $\td^{2+\ve}x$ at $\ve=0$.
Stated differently, all consistent definitions of ``$\td^{2+\ve}x$'' away from $\ve=0$ that one might come up with are
equivalent. The reason for that is the following.

Any integration over a scalar function on a manifold involves a volume form, i.e.\ a nowhere vanishing $d$-form (or a
density in the non-orientable case), in order to define a measure. This volume form is given by $\dd x\sg$, where $\sg$
is the square root of the corresponding Gramian determinant. If an integral is to be evaluated, the unit vectors of the
underlying coordinate system are inserted into the volume form. Since, for any $d$, these unit vectors produce a factor
of $1$ when inserted into $\dd x$, we see that it is the remaining part of the volume element that contains its
complete $d$-dependence, namely $\sg$. In particular, $\sg$ carries the canonical dimension of the volume
element.\footnote{Our conventions for the canonical mass dimensions are such that all coordinates are dimensionless,
$[x^\mu]=0$, while the metric components have $[g_\mn]=-2$, giving $\td s^2=g_\mn \td x^\mu\td x^\nu$ the canonical
dimension of an area, $[\td s^2]=-2$, whatever is the value of $d$. Hence $[\td x^\mu]=0$ and $[\sg]=-d$.

As a consequence,
the symbolic integration over the remaining ``fraction of a dimension'', $\td^\ve x$, is irrelevant even for the
dimension of $\mS_\ve[g]$.}

To summarize, for the evaluation of $\lim_{\ve\rightarrow 0} \frac{1}{\ve}\mS_\ve$ it is sufficient to consider the
$\ve$-depen\-dence of $\sg R$, while the integration can be seen as an integration over $\td^2x$. This
prescription can be considered our \emph{definition} for taking the $\ve$-limit in a well behaved way. Clearly,
the details of the domain of integration contribute some $\ve$-dependence, too. However, as we have seen in
\textbf{(1)} in equation \eqref{eqn:SVariation}, the first relevant non-constant, i.e.\ metric dependent, part of the
action comes from $\sg R$ alone, and any further $\ve$-dependent contributions would be of order $\ve^2$. This
makes clear that our argument is valid in the special case of an integral over $\sg R$, but not for arbitrary
integrands.

\medskip
\noindent
\textbf{(3) Comment and comparison with related work}. As an aside we note that in ref.\ \cite{MR93} it is argued that
the irrelevant divergent term in \eqref{eq:expansion} can be made vanish by subtracting the term $\frac{1}{\ve}
\int\dd x\sqrt{\tilde{g}}\,\tilde{R}$ from $\frac{1}{\ve}\int\dd x\sg \,R$ where the metric $\tilde{g}_\mn$ is assumed
to be $g_\mn$-dependent but chosen in such a way that the resulting field equations for $g_\mn$ do not change when $d$
approaches $2$. That means, the $g_\mn$-variation of the subtracted term (and, in turn its variation w.r.t.\
$\tilde{g}$) must vanish for $d\rightarrow 2$, leading to the requirement $\lim_{\ve\rightarrow 0} \big(\frac{1}{\ve}
\tilde{G}_\mn\big) = 0$ for the corresponding Einstein tensor.
This subtraction term would cancel the $\ve$-pole in \eqref{eq:expansion}. It is assumed in \cite{MR93} that such a
term exists for some metric $\tilde{g}_\mn$ which is conformally related to $g_\mn$. However, it remains unclear if
this is possible at all. According to the above argument in \textbf{(1)} we would rather expect $\frac{1}{\ve}
\tilde{G}_\mn$ to remain finite in the limit $\ve\rightarrow 0$.

Unlike ref.\ \cite{MR93}, we do not need to subtract further $g_\mn$-dependent terms from the action here, and our
discussion is valid for all metrics.

\subsection{Establishing the 2D limit}
\label{sec:EpsilonLimit}

Next we determine the first relevant order of the Taylor series, providing the basis for our main statements. Let us
define the $\ve$-dependent action functional
\begin{equation}
 \Ye[g] \equiv \frac{1}{\ve}\int\dex\sg\, R\, - \frac{4\pi\chi}{\ve}\,.
\label{eq:Ye}
\end{equation}
Here $\chi$ denotes again the metric independent Euler characteristic defined in strictly $2$ dimensions. Corresponding
to the arguments of section \ref{sec:genRem}, $\Ye$ is well defined in the limit $\ve\rightarrow 0$ since it is
of order $\ve^0$. Therefore, $Y[g]$ defined by
\begin{equation}
 Y[g] \equiv \lim_{\ve\rightarrow 0} \Ye[g]
\label{eq:YDef}
\end{equation}
is a finite functional.

To expand the integral in \eqref{eq:Ye} in powers of $\ve$ we make use of the general transformation law of
$\int\dd x\sg R$ under Weyl rescalings $g_\mn=\e^{2\sigma}\hg_\mn$, given by equation \eqref{eq:EHexpanded} in the appendix.
This yields
\begin{equation}
 \begin{split}
  \Ye[g] &= \frac{1}{\ve}\int\dex\shg\,\e^{\ve\sigma}\left[\hR+(1+\ve)\ve
    \big(\hD_\mu\sigma\big) \big(\hD^\mu\sigma\big)\right] - \frac{4\pi\chi}{\ve}\\
  &= \frac{1}{\ve}\int\dex\shg\, \hR - \frac{4\pi\chi}{\ve} 
    + \int\td^2 x\shg\big(\hR\sigma+\hD_\mu\sigma\hD^\mu\sigma \big) + \mO(\ve).
 \end{split}
\label{eq:YeExp1}
\end{equation}
We observe that the first two terms of the second line of \eqref{eq:YeExp1} can be combined into $\Ye[\hg]$.
Furthermore, the terms involving the parameter of the Weyl transformation, $\sigma$, are seen to agree with the
definition in \eqref{eq:DeltaIDef} and can be written as $\int\td^2x\shg\big[\hD_\mu\sigma\hD^\mu\sigma+\hR\sigma\big]
\equiv 2\, \Delta I[\sigma;\hg]$.
This, in turn, can be expressed by means of the (normalized) induced gravity functional \cite{P81}, defined
by\footnote{If the scalar Laplacian $\Box$ has zero modes, then $\Box^{-1}$ is defined as the inverse of $\Box$ on the
orthogonal complement to its kernel, that is, before $\Box^{-1}$ acts on a function it implicitly projects onto nonzero
modes. For the arguments presented in this section we may assume that $\Box$ does not have any zero modes, although a
careful analysis shows that the inclusion of zero modes does not change our main results (see detailed discussion in
appendix \ref{app:Weyl}, in particular section \ref{app:Zero}).}
\begin{equation}
 I[g] \equiv \int \td^2 x \sg\, R\, \Box^{-1}R \,.
\label{eq:IndGrav}
\end{equation}
As shown in appendix \ref{app:Weyl}, the change of $I$ under a finite Weyl transformation of the metric in its argument
equals precisely $-8\,\Delta I$ which therefore has the interpretation of a Wess--Zumino term, a $1$-cocycle related to
the Abelian group of Weyl transformations \cite{MM01}:\footnote{As a consequence of identity \eqref{eq:decomp1}, the
Liouville action \eqref{eq:LiouvilleAction} can be rewritten as $\GL[\phi;\hg]=\frac{a_1}{4}I[\e^{2\phi}\hg]+
\frac{1}{2}a_1 a_2\int\td^2x\sqrt{\det(\e^{2\phi}\hg)} - \frac{a_1}{4} I[\hg]$. Note that the first two terms on the
RHS of this equation depend on $\phi$ and $\hg_\mn$ only in the combination $\e^{2\phi}\hg_\mn = g_\mn$.}
\begin{equation}
 I[\e^{2\sigma}\hg] - I[\hg] = -8\,\Delta I[\sigma;\hg] \,.
\label{eq:decomp1}
\end{equation}
Inserting \eqref{eq:decomp1} into \eqref{eq:YeExp1} leads to
\begin{equation}
 \begin{split}
\Ye[g] = \Ye[\hg]+2\,\Delta I[\sigma;\hg] +\mO(\ve) 
  = \Ye[\hg] +\frac{1}{4}I[\hg]-\frac{1}{4}I[g] +\mO(\ve).
 \end{split}
\end{equation}
Rearranging terms and taking the limit $\ve\rightarrow 0$ results in the important identity
\begin{equation}
 Y[g]+\frac{1}{4}I[g]=Y[\hg]+\frac{1}{4}I[\hg].
\label{eq:YplusI}
\end{equation}
Note that the LHS of eq.\ \eqref{eq:YplusI} depends on the full metric $g=\e^{2\sigma}\hg$ while the RHS depends on
$\hg$ only.

For the further analysis it is convenient to introduce the functional
\begin{equation}
 F[g] \equiv Y[g] + \frac{1}{4} I[g].
\label{eq:FDef}
\end{equation}
By construction $F$ has the following properties:
{%
\renewcommand{\theenumi}{(\roman{enumi})}%
\setlist{nolistsep}
\begin{enumerate}
 \item It is diffeomorphism invariant since it has been constructed from diffeomorphism invariant objects only.
 \item It is a functional in $d=2$ precisely since the $\ve$-limit has already been taken.
 \item It is insensitive to the conformal factor of its argument since from eq.\ \eqref{eq:YplusI} follows
 Weyl invariance:
 \begin{equation}
  F[\e^{2\sigma}\hg] = F[\hg].
 \end{equation}
\end{enumerate}%
}%
\noindent
Thanks to our preparations in section \ref{sec:ConfGauge} we can conclude immediately that $F$ is constant apart from
a remaining dependence on some moduli $\{\tau\}$ possibly. Here it is crucial that the moduli are \emph{global}
parameters of purely \emph{topological} origin. They are insensitive to the local properties of the metric, in
particular they do not depend on a spacetime point. These arguments show that the \emph{functional} $F[g]$ becomes a
\emph{function} of the moduli, say $C\big(\{\tau\}\big)$. The precise dependence of $F$ on these moduli is irrelevant
for the present discussion since they encode only topological information. We thus have
\begin{equation}
 F[g] = C\big(\{\tau\}\big) ,
\end{equation}
i.e.\ $F$ is a metric independent constant functional, up to topological terms.

\bigskip
For the functional $Y[g]$ defined in eq.\ \eqref{eq:YDef} we obtain, using eq.\ \eqref{eq:FDef},
\begin{equation}
 Y[g]=-\frac{1}{4}I[g]+C\big(\{\tau\}\big) \, ,
\end{equation}
which leads to our final result:
\begin{equation}[b]
 \frac{1}{\ve}\int\td^{2+\ve}x\sg\, R= -\frac{1}{4}\int\td^2x\sg\,R\,\Box^{-1}R +\frac{4\pi\chi}{\ve}
  +C\big(\{\tau\}\big)+\mO(\ve).
\label{eq:LimitResult}
\end{equation}
The terms $4\pi\chi/\ve$ and $C\big(\{\tau\}\big)$ are topology dependent but independent of the local properties of
the metric, and thus they may be considered irrelevant for most purposes.

So we have established that the limit $d\rightarrow 2$ of the Einstein--Hilbert action equals precisely the induced gravity
action up to topological terms. Clearly the most remarkable aspect of this limiting procedure is that it leads from a local
to a non-local action.

A similar mechanism has been discussed earlier in the framework of dimensional regularization \cite{MM01}. The result
\eqref{eq:LimitResult} is in agreement with the one of reference \cite{J06} where it has been obtained by means of a different
reasoning based on the introduction of a Weyl gauge potential.

\subsection[The full Einstein--Hilbert action for \texorpdfstring{$\ve\rightarrow 0$}{epsilon to zero}]
  {The full Einstein--Hilbert action for \texorpdfstring{\bm{$\ve\rightarrow 0$}}{epsilon to zero}}
\label{sec:LimitFullEH}

Including also the cosmological constant term, the Einstein--Hilbert truncation of the (gravitational part of
the) effective average action reads
\begin{equation}
 \Gamma_k^\text{grav}[g] = \frac{1}{16\pi G_k} \int \dd x \sg \,\big( -R + 2\Lambda_k \big),
\label{eq:EHTruncation}
\end{equation}
with the dimensionful Newton and cosmological constant, $G_k$ and $\Lambda_k$, respectively.

\medskip
\noindent
\textbf{(1)}
As we have mentioned already, RG studies in $d=2+\ve$ show that the $\beta$-functions of the dimensionless versions
of these couplings, $g_k\equiv k^{d-2}G_k$ and $\lambda_k\equiv k^{-2}\Lambda_k$, possess a nontrivial fixed point
which is proportional to $\ve$ \cite{R98} (see also \cite{W80,T77,B77,GKT78,CD78,KN90,JJ91,KKM93,NR13,N15,CD15,F15}),
\begin{equation}
 g_*\propto\ve\quad\text{and}\quad \lambda_*\propto\ve.
\end{equation}
Thus, at least in the vicinity of this non-Gaussian fixed point the dimensionful couplings are of the form
\begin{equation}
 G_k \equiv \ve\,\breve{G}_k \, ,\quad \Lambda_k \equiv \ve\,\breve{\Lambda}_k \,,
\end{equation}
where $\breve{G}_k$ and $\breve{\Lambda}_k$ are of order $\mO(\ve^0)$. Making use of eq.\ \eqref{eq:LimitResult} in the
limit $\ve\rightarrow 0$ we arrive at the $2$-dimensional effective average action
\begin{equation}[b]
 \Gamma_k^\text{grav,2D}[g] = \frac{1}{64\pi \breve{G}_k} \int\td^2x\sg\,R\,\Box^{-1}R
  + \frac{\breve{\Lambda}_k}{8\pi\breve{G}_k}\int\td^2x\sg
  + \text{top}.
\label{eq:Gamma}
\end{equation}
Here 'top' refers again to topology dependent terms which are insensitive to the local properties of the metric. The
result \eqref{eq:Gamma} is quite general; it holds for any RG trajectory provided that the couplings $G_k$ and
$\Lambda_k$ in $d=2+\ve$ are of first order in $\ve$.

As an aside we note that the topological terms in \eqref{eq:Gamma} include a contribution proportional to
$\int\!\sg\,R=4\pi\mku\chi$. Thus, eq.\ \eqref{eq:Gamma} contains the induced gravity action, a cosmological constant
term, and the $\chi$-term. These are precisely the terms that were included in the truncation ansatz in ref.\
\cite{CD15}. By contrast, in our approach they are not put in by hand through an ansatz, but they rather emerge as a
result from the Einstein--Hilbert action in the 2D limit.

\medskip
\noindent
\textbf{(2)}
If we want to consider $\Gamma_k$ exactly at the NGFP, we can insert the known fixed point values, where the one of
Newton's constant is given by $g_*=\ve/b$ according to eq.\ \eqref{eq:NGFPWithB}. The coefficient $b$ is independent of
the cutoff scheme underlying the computation. It depends, however, on the parametrization of the metric. In the linear
parametrization, $g_\mn=\gb_\mn+h_\mn$, it is given by \cite{R98,W80,T77,B77,KN90,JJ91,NR13,N15,CD15}\footnote{When the
running of the Gibbons--Hawking surface term instead of the pure Einstein--Hilbert action is computed, the result reads
$b=\frac{2}{3}(1-\ns)$ \cite{GKT78,CD78}. See ref.\ \cite{NR13} for a discussion.}
\begin{equation}
 b=\frac{2}{3}\big(19-\ns\big),
\end{equation}
while the exponential parametrization \cite{DN15}, $g_\mn=\gb_{\mu\rho}(\e^h)^\rho{}_\nu$, leads to
\cite{KKM93,N15,CD15,F15}
\begin{equation}
 b=\frac{2}{3}\big(25-\ns\big).
\label{eq:bExp}
\end{equation}
While we will argue in section \ref{sec:Emergence} that the two parametrizations might possibly describe different
universality classes, to make contact to the results known from $2$-dimensional conformal field theory the exponential
parametrization turns out to be more appropriate. Therefore, we will mostly state the results based on eq.\
\eqref{eq:bExp} in the following, although the analogues for the linear parametrization can be obtained simply by
replacing $25\rightarrow 19$. Using the definition \eqref{eq:IndGrav} and combining \eqref{eq:Gamma} with
\eqref{eq:bExp} we obtain the NGFP action
\begin{equation}[b]
 \Gamma_k^\text{grav,2D,NGFP}[g] = \frac{(25-\ns)}{96\pi} \,I[g]
  + \frac{(25-\ns)}{12\pi}\,k^2\breve{\lambda}_*\int\td^2x\sg
  + \text{top}\,,
\label{eq:GammaNGFP}
\end{equation}
where $\breve{\lambda}_*\equiv\lambda_*/\ve$ is cutoff dependent and thus left unspecified here. The actions
\eqref{eq:Gamma} and \eqref{eq:GammaNGFP} will be the subject of our discussion in section \ref{sec:NGFPCFT}.

\medskip
\noindent
\textbf{(3)}
Finally, let us briefly establish the connection with Liouville theory. For this purpose we separate the conformal
factor from the rest of the metric. Inserting
\begin{equation}
 g_\mn = \e^{2\phi}\hg_\mn
\label{eq:DefConfFac}
\end{equation}
into eq.\ \eqref{eq:Gamma} for $\Gamma_k^\text{grav,2D}[g]$ and using \eqref{eq:ItoDeltaI} and \eqref{eq:DeltaI} from
the appendix yields
\begin{equation}[b]
\begin{aligned}
 \Gamma_k^\text{grav,2D}[\phi;\hg] =\; &\frac{1}{64\pi\breve{G}_k}\int\td^2x\shg\,\hR\,\hB^{-1}\hR \\
  & -\frac{1}{16\pi\breve{G}_k}\int\td^2x\shg\,\Big[\hD_\mu\phi\,\hD^\mu\phi+\hR\phi-2\breve{\Lambda}_k \e^{2\phi}\Big]
  + \text{top} \,,
\end{aligned}
\end{equation}
where $\hg_\mn$ is a fixed reference metric for the topological sector (i.e.\ a point in moduli space) under
consideration. Hence, the effective average action for the conformal factor in precisely $2$ dimensions is nothing but
the Liouville action.

Of course, this is well known to happen if one starts from the induced gravity action, an object that lives already in
2D. It is quite remarkable and nontrivial, however, that \emph{Liouville theory can be regarded the limit of the higher
dimensional Einstein--Hilbert theory}. Note that this result is consistent with the discussions in
refs.\ \cite{MR93,GJ10} (cf.\ also \cite{LS94}).

\subsection{Aside: Is there a generalization to 4D?}
\label{sec:GenTo4D}

For the sake of completeness we would like to comment on a generalization of our results to $4$ dimensions. At first
sight there seems to be a remarkable similarity. Dimensional analysis suggests that the role of the $R$-term in the
Einstein--Hilbert action near $2$ dimensions is now played by curvature-square terms in $d=4+\ve$. The gravitational
part of the action assumes the form
$\Gamma_k[g]=\Gamma_k^\text{EH}+\int\td^{4+\ve}x\sg\left\{\frac{1}{a_k}E+\frac{1}{b_k}F+\frac{1}{c_k}R^2\right\}$
where $F\equiv C_{\mn\rs}C^{\mn\rs}$ is the square of the Weyl tensor, and
$E\equiv R_{\mn\rs}R^{\mn\rs}-4R_\mn R^\mn+R^2+\frac{d-4}{18}R^2$ gives rise to the Gauss--Bonnet--Euler topological
invariant when integrated over in $d=4$. Considerations of nontrivial cocycles of the Weyl group show that the
corresponding Wess--Zumino action in $d=4$ is generated by the $E$- and the $F$-term \cite{MM01}, analogous to the
generation of $\Delta I$ in sec.\ \ref{sec:EpsilonLimit} due to the $R$-term. It may thus be expected that there would
be a mechanism to take the 4D limit, similar to the one of sec.\ \ref{sec:EpsilonLimit} but now for $E$ and $F$ instead
of $R$, if the couplings $a_k$ and $b_k$ were of first order in $\ve$.

At one-loop level the $\beta$-functions in $d=4+\ve$ have indeed a fixed point with $a_*=\mO(\ve)$, $b_*=\mO(\ve)$ and
$c_*$ finite \cite{OP14}. There are, however, two crucial differences in comparison with the $2$-dimensional case:
(i) The term $\int\td^4 x\sg \,F$ is not a topological invariant, i.e.\ there is no appropriate subtraction analogous
to definition \eqref{eq:Ye}, and the limit $\ve\rightarrow 0$ remains problematic. (ii) Even if we managed to define
some 4D-functional similar to \eqref{eq:FDef} which is both diffeomorphism and Weyl invariant, this would not be
sufficient to conclude that the functional is constant since in $d=4$ the space of metrics modulo
$\text{Diff}\times\text{Weyl}$-transformation is too large and cannot be classified in terms of topological
parameters. Roughly speaking, if we found a way to circumvent problem (i), the 4D limit of the above action computed
with our methods might lead to the same non-local action as found in \cite{MM01}, but this would not represent the
general 4D limit since the latter must certainly contain further terms that do not originate from a variation of the
conformal factor alone.
In summary, in spite of many similarities to the 2D case there seems to be no direct generalization of our
approach of computing a non-local limit action to $4$ spacetime dimensions. Nevertheless, we expect that the 4D fixed
point action contains non-local terms, too.

\section{The NGFP as a conformal field theory}
\label{sec:NGFPCFT}

We can summarize the previous sections by saying that every trajectory $k \mapsto (g_k,\lambda_k)\equiv (\bgk,\blk)\ve$,
i.e.\ every solution to the RG equations of the Einstein--Hilbert truncation in $2+\ve$ dimensions, induces the
following intrinsically $2$-dimensional running action:
\begin{equation}
 \Ggd[g] = \frac{1}{96\pi}\left(\frac{3}{2}\,\frac{1}{\bgk}\right)\left[ I[g]+8\blk\,k^2\int\td^2 x\sg \,\right].
\label{eq:GammaGravSummarized}
\end{equation}
In this section we discuss the main properties of this RG trajectory, in particular its fixed point.

\medskip
\noindent
\textbf{(1) The fixed point functional}. Strictly speaking, the theory space under consideration comprises functionals
which depend on the \emph{dimensionless} metric $\tg_\mn\equiv k^2 g_\mn$. For any average action $\Gamma_k[g]$ we
define its analog in the dimensionless setting by $\mA_k[\tg]\equiv\Gamma_k[\tg\mku k^2]$. Thus, equation
\eqref{eq:GammaGravSummarized} translates into
\begin{equation}
 \mA_k[\tg] = \frac{1}{96\pi}\left(\frac{3}{2}\,\frac{1}{\bgk}\right)\left[ I[\tg]+8\blk \int\td^2 x\stg \,\right].
\end{equation}
It is this functional that becomes strictly constant at the NGFP: $\mA_k\rightarrow \mA_*\mku$, with
\begin{equation}
 \mA_*[\tg] = \frac{1}{96\pi}\left(\frac{3}{2}\,\frac{1}{\bgs}\right)\left[ I[\tg]+8\bls \int\td^2 x\stg \,\right].
\label{eq:AStar}
\end{equation}
For the exponential field parametrization we find the fixed point functional
\begin{equation}[b]
 \mA_*[\tg] = \frac{(25-\ns)}{96\pi} \int\td^2 x\stg\,\Big(\tR\,\tB^{-1}\tR+8\bls\Big) .
\end{equation}
Here and in the following we usually present the results for the exponential parametrization. The corresponding
formulae for the linear parametrization can be obtained by replacing $(25-\ns)\rightarrow(19-\ns)$. See also section
\ref{sec:Emergence} for a discussion of the different metric parametrizations.

While the NGFP is really a point in the space of $\mA$-functionals, it is an entire \emph{line}, parametrized by $k$,
in the more familiar dimensionful language of the $\Gamma_k$'s. Let us refer to the constant map $k\mapsto(g_*,
\lambda_*)$ $\forall\, k\in[0,\infty)$ as the ``\emph{FP trajectory}''. Moving on this trajectory, the system is never
driven away from the fixed point. According to eq.\ \eqref{eq:GammaNGFP}, it is described by the following EAA:
\begin{equation}[b]
 \Gamma_k^\text{grav,2D,NGFP}[g] = \frac{(25-\ns)}{96\pi} \left[ I[g]+8\mku\bls\, k^2\int\td^2 x\sg \,\right].
\label{eq:GgAtNGFP}
\end{equation}
As always in the EAA framework, the EAA at $k=0$ equals the standard effective action, $\Gamma =
\lim_{k\rightarrow 0}\Gamma_k$. So, letting $k=0$ in \eqref{eq:GgAtNGFP}, we conclude that the ordinary effective action
related to the FP trajectory has vanishing ``renormalized'' cosmological constant and reads
\begin{equation}
 \Gamma^\text{grav,2D,NGFP}[g] = \frac{(25-\ns)}{96\pi} \int\td^2 x\sg\,R\,\Box^{-1} R\,.
\label{eq:GgZeroAtNGFP}
\end{equation}

\medskip
\noindent
\textbf{(2) The 2D stress tensor}. Differentiating $\Ggd$ of equation \eqref{eq:GammaGravSummarized} with respect to the
metric leads to the following energy-momentum tensor in the gravitational sector \cite{CR89}:
\begin{equation}
\begin{split}
 T_\mn^\text{grav}[g] = \frac{1}{96\pi}\left(\frac{3}{2}\,\frac{1}{\bgk}\right)\bigg[  g_\mn\, D_\rho\big(\Box^{-1}R\big)
 D^\rho\big(\Box^{-1}R\big) - 2\, D_\mu\big(\Box^{-1}R\big) D_\nu\big(\Box^{-1}R\big) & \\
 + 4\, D_\mu D_\nu\big(\Box^{-1}R\big) -4\, g_\mn R + 8\, \blk\,k^2 g_\mn & \bigg].
\end{split}
\label{eq:Tgrav}
\end{equation}
It is easy to see that taking the trace of this tensor yields $\Theta_k[g]=\left(\frac{3}{2}\,\frac{1}{\bgk}\right)
\frac{1}{24\pi}\big[-R+4\blk\,k^2\big]$ which, as it should be, agrees with the result from the Einstein--Hilbert action in
$d>2$, see equations \eqref{eq:Theta2} and \eqref{eq:Theta3}.\footnote{Note that in string theory or conformal field theory
one would usually redefine the stress tensor and employ $T_\mn'\equiv T_\mn-\frac{1}{2}g_\mn\Theta$ which is traceless at
the expense of not being conserved. It is the modes of $T_\mn'$ that satisfy a Virasoro algebra whose central extension
keeps track of the anomaly coefficient then.} As for the non-trace parts of $T_\mn^\text{grav}$, the comparatively
complicated non-local structures in \eqref{eq:Tgrav} can be seen as the 2D replacement of the Einstein tensor in
\eqref{eq:EHEMTensor}.

In absence of matter ($\GM=0$) the tadpole equation \eqref{eq:Tadpole} boils down to $T_\mn^\text{grav}[\gsc]=0$ with
the above stress tensor. Hence, self-consistent backgrounds have a constant (but $k$-dependent) Ricci scalar:
\begin{equation}
 \Theta_k[\gsc]=0 \quad\Leftrightarrow\quad R\big(\gsc\big) = 4 \blk\,k^2 \,.
\label{eq:Rsc}
\end{equation}
In terms of the dimensionless metric, $R\big(\tilde{\bar{g}}_k^\text{sc}\big)=4\blk$, in this case.

\medskip
\noindent
\textbf{(3) Intermezzo on induced gravity}. As a preparation for the subsequent discussion consider an arbitrary conformal
field theory on flat Euclidean space, having central charge $c_\mS$, and couple this theory to a gravitational background
field $g_\mn$, comprised in an action functional $\mS[g]$. Then the resulting (symmetric, conserved) stress tensor,
\begin{equation}
 T^{(\mS)}[g]^\mn \equiv \frac{2}{\sg}\frac{\delta\mS[g]}{\delta g_\mn} \,,
\label{eq:StressTensorGeneral}
\end{equation}
will acquire a nonzero trace in curved spacetimes, of the form
\begin{equation}
 g_\mn\, T^{(\mS)}[g]^\mn = -c_\mS\, \frac{1}{24\pi}R+\text{const} \,,
\label{eq:TCentralCharge}
\end{equation}
where ``const'' is due to a cosmological constant possibly.

\medskip
\noindent
\textbf{(3a)} Above $\mS[g]$ can stand for either a classical or an effective action.
In the first case, $\mS[g]$ might result from a CFT of fields $\chi^I$ governed by an action $S[\chi,g]$ upon solving the
equations of motion for $\chi$, and substituting the solution $\chi_\text{sol}(g)$ back into the action:
$\mS[g] = S[\chi_\text{sol}(g),g]$. If $c_\mS \neq 0$ then the system displays a ``classical anomaly'', and Liouville
theory is the prime example \cite{D91,GM92,N04,AADZ94}.

In the ``effective'' case, $\mS[g]$ could be the induced gravity action $S_\text{ind}[g]$ which we obtain from $S[\chi,g]$
by integrating out the fields $\chi^I$ quantum mechanically:
\begin{equation}
 \e^{-S_\text{ind}[g]} = \int\mD\chi^I\,\e^{-S[\chi,g]}\,.
\end{equation}
Then $S_\text{ind}$ is proportional to the central charge $c_\mS$,
\begin{equation}
 S_\text{ind}[g] = +\frac{c_\mS}{96\pi} I[g] + \cdots \,,
\label{eq:Sind}
\end{equation}
and by \eqref{eq:StressTensorGeneral} the action $S_\text{ind}$ gives rise to a stress tensor whose trace is precisely
of the form \eqref{eq:TCentralCharge}. (The dots represent a cosmological constant term.)

\medskip
\noindent
\textbf{(3b)} It is important to observe that the functional $I[g]$ is \emph{negative}, i.e.\ for any metric $g$ we
have $\int\td^2 x\sg\, R\,\Box^{-1}R <0$ . (Recall that $\Box^{-1}$ acts only on nonzero modes while it ``projects
away'' the zero modes. Since $-\Box$ is non-negative, we conclude that $-\Box^{-1}$ has a strictly positive spectrum.)
Leaving the cosmological constant term in \eqref{eq:Sind} aside, this entails that for a positive central charge
$c_\mS>0$ the (non-cosmological part of the) induced gravity action is negative, $S_\text{ind}[g]<0$.

The implications are particularly obvious in the conformal parametrization $g=\e^{2\phi} \hg$, yielding
\begin{equation}
 S_\text{ind}[\phi;\hg] = -\frac{c_\mS}{24\pi} \int\td^2 x\shg\Big(\hD_\mu\phi \hD^\mu\phi + \hR\phi\Big)
 + \frac{c_\mS}{96\pi} I[\hg] + \cdots \,.
\label{eq:instab}
\end{equation}
When $c_\mS$ is positive, the field $\phi$ is unstable, it has a ``wrong sign'' kinetic term. Stated differently,
\emph{integrating out unitary conformal matter induces an unstable conformal factor of the emergent spacetime metric.}

The 4D Einstein--Hilbert action is well known to suffer from the same conformal
factor instability, that is, a negative kinetic term for $\phi$ if the overall prefactor of $\int\!\sg R$ is adjusted
in such a way the concomitant kinetic term for the transverse-traceless (TT) metric fluctuations comes out positive,
as this befits propagating physical modes. Irrespective of all questions about the conventions in which the equations
are written down, the crucial signs are always such that
\begin{equation}
 c_\mS>0 \quad \stackrel{\mathclap{d=2}}{\Longleftrightarrow} \quad \phi \text{ unstable} \quad
 \stackrel{\mathclap{d>3}}{\Longleftrightarrow} \quad h_\mn^\text{TT} \text{ stable}.
\end{equation}
We shall come back to this point in a moment.

\medskip
\noindent
\textbf{(4) Central charge of the NGFP}. The fixed point action $\mA_*^\text{grav,2D}$ given by \eqref{eq:AStar}
describes a conformal field theory with central charge
\begin{equation}
 \cgr = \begin{cases} \;25 - \ns , \qquad\text{exponential parametrization}\\
 \;19-\ns , \qquad\text{linear parametrization.} \end{cases}
\label{eq:CentChParams}
\end{equation}
This follows by observing that for the two field parametrizations, directly at the NGFP, the trace of the stress tensor
is given by
\begin{equation}[b]
 \Theta_k[g] = \frac{1}{24\pi}\Big(-R+4\bls k^2\Big)\times\begin{cases}\;25-\ns\qquad\text{(exp.)}\\
 \;19-\ns\qquad\text{(lin.)\,.}\end{cases}
\label{eq:ThetaParams}
\end{equation}
Applying the rule \eqref{eq:TCentralCharge} to eq.\ \eqref{eq:ThetaParams}, we see indeed that, first, the fixed point
theory is a CFT, and second, its central charge is given by \eqref{eq:CentChParams}.\footnote{Reading off the central
charge according to \eqref{eq:TCentralCharge} and \eqref{eq:Sind} is consistent with refs.\ \cite{CDP14,CD15} where the
relation between the central charge and the $\beta$-function of Newton's constant is discussed in the FRG framework,
implying a relation between $\cgr$ and $g_*$.}

According to eq.\ \eqref{eq:GgAtNGFP}, the EAA related to the FP trajectory, $\Gamma_k^\text{grav,2D,NGFP}$, happens
to have exactly the structure of the induced gravity action \eqref{eq:Sind} with the corresponding central charge, for
all values of the scale parameter.

At the $k=0$ endpoint of this trajectory, the dimensionful cosmological constant $\bLk=\bls k^2$ runs to zero without
any further ado, and $\Gamma_{k\rightarrow 0}^\text{grav,2D,NGFP}$ becomes the standard effective action
\eqref{eq:GgZeroAtNGFP}. At this endpoint, by eq.\ \eqref{eq:Rsc}, self-consistent backgrounds have vanishing curvature
in the absence of matter: $R(\gb_{k=0}^\text{sc})=0$. Therefore, we have indeed inferred a central charge pertaining to
flat space by comparing \eqref{eq:ThetaParams} to \eqref{eq:TCentralCharge}.

\medskip
\noindent
\textbf{(5) Auxiliary ``matter'' CFTs}. Since the 2D gravitational fixed point action is of the induced gravity type,
we can, if we wish to, introduce a conformal matter field theory which induces it when the fluctuations of those
auxiliary matter degrees of freedom are integrated out (although such auxiliary fields are not required by our
formalism). Denoting the corresponding fields by $\chi^I$ again, and their ($k$ independent) action by
$S_\text{aux}[\chi;g]$, we have then
\begin{equation}[b]
 \e^{-\Gamma_k^\text{grav,2D,NGFP}[g]} \equiv \int\mD\chi\;\e^{-S_\text{aux}[\chi;g]}\cdot\e^{-N[g]} \; .
\label{eq:GammaSAux}
\end{equation}
Here $N[g]\propto\int\td^2 x\sg$ is a inessential correction term to make sure that also the nonuniversal cosmological
constant terms agree on both sides of \eqref{eq:GammaSAux}; it depends on the precise definition of the functional
integral.

Clearly the auxiliary matter CFT can be chosen in many different ways, the only constraint is that it must have the
correct central charge, $c_\text{aux}=\cgr$, that is, $c_\text{aux}=25-\ns$ or
$c_\text{aux}=19-\ns$, respectively. Here are two examples of auxiliary CFTs:

\medskip
\noindent
\textbf{(5a) Minimally coupled scalars}.
For $c_\text{aux}>0$ the simplest choice is a multiplet of minimally coupled scalars $\chi^I(x)$,
$I=1,\cdots,c_\text{aux}$. These auxiliary fields may not be confused with the physical matter fields $A^i(x)$,
$i=1,\cdots,\ns$. The $\chi$'s and $A$'s have nothing to do with each other except that their respective numbers must
add up to $25$ (or to $19$).

\medskip
\noindent
\textbf{(5b) Feigin--Fuks theory}. The induced gravity action $I[g]$ being a non-local functional, it is natural to
introduce one, or several fields in addition to the metric that render the action local. The minimal way to achieve
this is by means of a single scalar field, $B(x)$, as in Feigin--Fuks theory \cite{CT74}, which has a nonminimal
coupling to the metric. Consider the following local action, invariant under general coordinate transformations applied
to $g_\mn$ and $B$:
\begin{equation}
 I_\text{loc}[g,B] \equiv \int\td^2 x\sg\,\big(D_\mu B\,D^\mu B + 2\mku R\mku B\big)\,.
\label{eq:FFAction}
\end{equation}
The equation of motion $\delta I_\text{loc}/\delta B = -2\sg\,(\Box B-R) = 0$ is solved by $B=B(g)\equiv\Box^{-1}R$
which when substituted into $I_\text{loc}$ reproduces precisely the non-local form of the induced gravity action:
$I_\text{loc}[g,B(g)] = \int\sg\,R\,\Box^{-1}R\equiv I[g]$.

As $I_\text{loc}$ is quadratic in $B$, the same trick works
also quantum mechanically when we perform the Gaussian integration over $B$ rather than solve its field equation. Hence,
the exponentiated $\Gamma_k^\text{grav,2D,NGFP}$ has the representation
\begin{equation}
 \e^{-\frac{(25-\ns)}{96\pi}\, I[g] + \cdots} = \int\mD B\; \e^{-\frac{(24-\ns)}{96\pi}\int\td^2 x\sg\,(
  D_\mu B\,D^\mu B + 2\mku R\mku B + \cdots ) }
\end{equation}
Here again the dots stand for a cosmological constant which depends on the precise definition of the functional measure
$\mD B$. It is well known that thanks to the $R\mku B$-term the CFT of the $B$-field (in the limit $g_\mn\rightarrow
\delta_\mn$) has a shifted central charge \cite{FMS97,M10}; in the present case this reproduces the values
\eqref{eq:CentChParams}.

So the conclusion is that while the fixed point action is a non-local functional $\propto\int\sg\,R\,\Box^{-1}R$ in
terms of the metric alone, one may introduce additional fields such that the same physics is described by a local
(concretely, second-derivative) action. In particular, $\Gamma_k^\text{grav,2D,NGFP}$ and the local functional
\begin{equation}
 \Gamma_k^\text{loc}[g,B]\equiv \frac{(24-\ns)}{96\pi}\int\td^2 x\sg\,\big(D_\mu B\,D^\mu B + 2\mku R\mku B+\cdots\big)
\end{equation}
are fully equivalent, even quantum mechanically.

\medskip
\noindent
\textbf{(6) Positivity in the gravitational sector}. Pure quantum gravity ($\ns=0$) and quantum gravity coupled to
less than $25$ (or $19$) scalars are governed by a \emph{fixed point CFT with a positive central charge}.

Clearly, this is good news concerning the pressing issue of unitarity (Hilbert space positivity) in asymptotically safe
gravity. The theories with with $\cgr>0$, continued to Lorentzian signature, do indeed admit a quantum
mechanical interpretation and have a state space which is a Hilbert space in the mathematical sense (no negative norm
states), supporting a unitary representation of the Virasoro algebra \cite{S08}.

\medskip
\noindent
\textbf{(6a) Schwinger term}. Leaving the analytic continuation to the Lorentzian world aside, it is interesting to
note that already in Euclidean space the simple-looking induced gravity action ``knows'' about the fact that
$c_\text{grav}^\text{NGFP}<0$ would create a problem for the probability interpretation. By taking two functional
derivatives of the standard effective action \eqref{eq:GgZeroAtNGFP} we can compute the $2$-point function
$\langle\, T_\mn^\text{grav}(x) T_{\rho\sigma}^\text{grav}(y)\,\rangle$ and, in particular, its contracted form
$\langle\,\Theta_0(x)\Theta_0(y)\, \rangle$. Setting thereafter $g_\mn=\delta_\mn$, which as we saw is a
self-consistent background, and choosing a suitable coordinate chart, we obtain the following Schwinger term:
\begin{equation}[b]
 \langle\,\Theta_0(x)\Theta_0(y)\,\rangle = -\frac{\cgr}{12\pi}\, \p^\mu\p_\mu \delta(x-y)\,.
\label{eq:SchwingerTerm}
\end{equation}
Let us smear $\Theta_0$ with a real valued test function $f$ that vanishes at the boundary and outside of the chart
region, or, in the case where the chart is the entire Euclidean plane, falls off rapidly at infinity:\footnote{Note
that in the latter case the function $f$ has support on the entire Euclidean plane, hence we are clearly not testing
Osterwalder--Schrader \cite{OS73} reflection positivity here \cite{S93,GJ87}.}
$\Theta_0[f]\equiv\int\td^2 x\, f(x)\Theta_0(x)$. Then, applying $\int\td^2 x\,\td^2 y\,f(x)f(y)\cdots$ to both sides
of \eqref{eq:SchwingerTerm}, we find after an integration by parts:
\begin{equation}
 \langle\,\Theta_0[f]^2\,\rangle = 
  + \, \cgr\,\frac{1}{12\pi}\int\td^2 x\,(\p_\mu f)\delta^\mn(\p_\nu f)\,.
\label{eq:ThetaSquared}
\end{equation}
Since the integral on the RHS of \eqref{eq:ThetaSquared} is manifestly positive, we conclude that if
$\cgr<0$ the expectation value of the square $\Theta_0[f]^2$ is negative. Clearly this would be
problematic already in the context of statistical mechanics (at least with real field variables).

\medskip
\noindent
\textbf{(6b) Induced gravity approach in 4D: a comparison}. Note that one can extract the central charge from the
Schwinger term by performing an integral $\int\td^2 x\, x^2(\cdots)$ over both sides of eq.\ \eqref{eq:SchwingerTerm}.
Since Newton's constant is dimensionless in 2D, and $\breve{G}^{-1}=\bgs^{-1}=b=\frac{2}{3}\,\cgr$, this leads to the
following integral representation for the Newton constant belonging to the 2D world governed by the FP trajectory
\cite{A82}:
\begin{equation}[b]
 \breve{G}^{-1} = -2\pi\int\td^2 x\; x^2\langle\,\Theta_0(x)\Theta_0(0)\,\rangle \,.
\end{equation}
It is interesting to note that this representation is of precisely the same form as the relations that had been derived
long ago within the induced gravity approach in 4D, the hope being that ultimately one should be able to compute its RHS
from a matter field theory, assumed known, the Standard Model, say, and would then predict the value of Newton's
constant in terms of matter-related constants of Nature.

For a review and a discussion of the inherent difficulties we refer to \cite{A82}. We see that, in a sense, Asymptotic
Safety was successful in making this scenario work, producing a positive Newton constant in particular, but with one
key difference: The underlying matter field theory, here the `aux' system, is no longer an arbitrary external input,
but is chosen so as to reproduce the NGFP action, an object computed from first principles.

\medskip
\noindent
\textbf{(7) Complete vs.\ gauge invariant fixed point functional}. So far we mainly focused on the gravitational part
of the NGFP functional. The complete EAA, namely $\Gamma_k = \Gg +\GM +\Gamma_k^\text{gf} +\Gamma_k^\text{gh}$ contains
matter, gauge fixing and ghost terms in addition. But since the present truncation neglects the running of the latter
three parts, they may be considered always at their respective fixed point. Also, they have an obvious interpretation
in 2D exactly. Furthermore, our truncation assumes that neither $\Gg$ nor $\GM$ as given in \eqref{eq:matter} has an
``extra'' $\gb$-dependence.

As a result, the sum of gravity and matter (`GM') contributions,
\begin{equation}
 \Gamma_k^\text{GM,2D}[g,A] \equiv
  \Ggd[g]+\frac{1}{2}\sum\limits_{i=1}^{\ns}\int\td^2 x\sg\;g^\mn\p_\mu A^i\p_\nu A^i \,,
\end{equation}
enjoys both Background Independence, here meaning literally independence of the background metric, and Gauge Invariance,
i.e.\ it does not change under diffeomorphisms applied to $g_\mn$ and $A^i$.

Thanks to the second property\footnote{Which might not be realized in more complicated truncations!},
we may adopt the point of view that it is actually the gauge invariant functional $\Gamma_k^\text{GM,2D}$ alone which
contains all information of interest and was thus ``handed over'' alone from the higher dimensional Einstein--Hilbert
world to the intrinsically $2$-dimensional induced gravity setting.
Therefore, if in 2D the necessity of gauge fixing arises, we can in principle pick a new gauge, different from
the one employed in $d>2$ for the computation of the $\beta$-functions.\footnote{This could not be
done if one wants to combine loop or RG calculations from $d>2$ with others done in $d=2$ exactly. However, in the
present paper all dynamical calculations are done in $d>2$, i.e.\ before the 2D limit is taken}

\medskip
\noindent
\textbf{(8) Unitarity vs.\ stability: the conformal factor ``problem''}.
Next we take advantage of the particularly convenient conformal gauge available in strictly $2$ dimensions (cf.\
section \ref{sec:ConfGauge}), and evaluate $\GgdN[g]$ as given explicitly by eq.\ \eqref{eq:GgAtNGFP} for metrics of
the special form $g_\mn = \hg_\mn\,\e^{2\phi}$. The result is a Liouville action as before in eqs.\
\eqref{eq:LiouvilleAction}, \eqref{eq:GravToLiou}, this time without any undetermined piece such as $U_k[\hg]$,
however:
\begin{align}
 \GgdN[\hg\mku\e^{2\phi}] &\equiv \frac{\cgr}{96\pi}\,I[\hg] + \GL[\phi;\hg]\,,
\label{eq:GammaGravILiou}\\
 \GL[\phi;\hg] &= \frac{\cgr}{12\pi}\int\td^2 x\shg\,\left\{-\frac{1}{2}\,\hD_\mu\phi\,\hD^\mu\phi-\frac{1}{2}\hR\phi
 +\bls\,k^2\mku\e^{2\phi} \right\} \,.
\label{eq:LiouvilleNGFP}
\end{align}
Since $\cgr=25-\ns$ (or $\cgr=19-\ns$ with the linear parametrization), we observe that for pure gravity, and gravity
interacting with not too many matter fields, the conformal factor has a ``wrong sign'' kinetic term that might seem
to indicate an instability at first sight. If we think of the fixed point action as induced by some auxiliary CFT
with central charge $c_\text{aux}=\cgr=25-\ns>0$, we see that this is exactly the correlation mentioned in paragraph
\textbf{(3b)} above: bona fide unitary CFTs generate ``wrong sign'' kinetic terms for the conformal factor.

We emphasize that the unstable $\phi$-action is neither unexpected, nor ``wrong'' from the physics point of view,
nor in contradiction with the positive central charge of the fixed point CFT. Let us discuss these issues in turn
now.

\medskip
\noindent
\textbf{(8a) The importance of Gauss' law}. Recall the standard count of gravitational degrees of freedom in
Einstein--Hilbert gravity: In $d$ dimensions, the symmetric tensor $g_\mn$ contains $\frac{1}{2}d(d+1)$ unknown
functions which we try to determine from the $\frac{1}{2}d(d+1)$ field equations $G_\mn=\cdots\,$. Those are not
independent, but subject to $d$ Bianchi identities. Moreover, we need to impose $d$ coordinate conditions due to
diffeomorphism invariance. This leaves us with $\NEH(d)\equiv\frac{1}{2}d(d+1)-d-d=\frac{1}{2}d(d-3)$ gravitational
degrees of freedom, meaning that by solving the Cauchy problem for $g_\mn$ we can predict the time evolution of
$\NEH(d)$ functions that, (i), are related to ``physical '' (i.e.\ gauge invariant) properties of space, (ii), are
algebraically independent among themselves, and (iii), are \emph{independent of the functions describing the evolution
of matter}.

With $\NEH(4)=2$ we thus recover the gravitational waves of 4D General Relativity, having precisely $2$
polarization states. Similarly, $\NEH(3)=0$ tells us that there can be no gravitational waves in $3$
dimensions since all independent, gauge invariant properties described by the metric can be inferred already from
the matter evolution. No extra initial conditions can, or must, be imposed.

Finally $\NEH(2)=-1$ seems to suggest that ``gravity has $-1$ degree of freedom in $2$ dimensions''. Strange as it
might sound, the meaning of this result is quite clear: the quantum metric with its ghosts removes one degree of
freedom from the matter system.
If, in absence of gravity, the Cauchy problem of the matter system has a unique solution after specifying $N_\text{M}$
initial conditions, then this number gets reduced to $N_\text{M}-1$ by coupling the system to gravity.

Quantum mechanically, on a state space with an indefinite metric, the removing of degrees of freedom happens upon
imposing ``Gauss law constraints'', or ``physical state conditions'' on the states. As a result, the potentially
dangerous negative-norm states due to the wrong sign of the kinetic term of $\phi$ are not part of the actual
(physical) Hilbert space. The latter can be built using matter operators alone, and it is smaller in fact than without
gravity.\footnote{See Polchinski \cite{P89} for a related discussion.}

The situation is analogous to Quantum Electrodynamics (QED) in Coulomb gauge, for example. The overall sign of the
Maxwell action $\propto F_\mn F^\mn$ is chosen such that the spatial components of $A_\mu$ have a positive kinetic term,
and so it is unavoidable that the time component $A_0$ has a negative one, like the conformal factor in
\eqref{eq:LiouvilleNGFP}. However, it is well known \cite{BD65} that the states  with negative $(\text{norm})^2$
generated by $A_0$ do not survive imposing Gauss' law $\bm{\nabla}\cdot\bm{E}=e\,\psi^\dagger\psi$ on the states. This
step indeed removes one degree of freedom since $A_0$ and $\rho_\text{em}\equiv e\,\psi^\dagger\psi$ get coupled
by an \emph{instantaneous} equation, $\bm{\nabla}^2 A_0(t,\bm{x})=-\rho_\text{em}(t,\bm{x})$.

\medskip
\noindent
\textbf{(8b) Instability and attractivity of classical gravity}. To avoid any misunderstanding we recall that in
constructing realistic 4D theories of gravity it would be quite absurd, at least in the Newtonian limit, to ``solve''
the problem of the conformal factor by manufacturing a positive kinetic term for it in some way. In taking the
classical limit of General Relativity this kinetic term descends essentially to the $\bm{\nabla}\vp_\text{N}\cdot
\bm{\nabla}\vp_\text{N}$-part of the classical Lagrangian governing Newton's potential $\vp_\text{N}$ and therefore
fixes the positive sign on the RHS of Poisson's equation, $\bm{\nabla}^2\vp_\text{N}=+4\pi\mku G\rho$. However, this
latter plus sign expresses nothing less than the universal attractivity of classical gravity, something we certainly
want to keep.

This simple example shows that the conformal factor instability is by no means an unmistakable sign for a
physical deficiency of the theory under consideration. The theory can be perfectly unitary if there are appropriate
Gauss' law-type constraints to cut out the negative norm states of the indefinite metric state space.

\medskip
\noindent
\textbf{(8c) Central charge in Liouville theory}. Finally, we must discuss a potential source of confusion concerning
the correct identification of the fixed point's central charge. Let us pretend that the Liouville action
$\GL[\phi;\hg]$ describes a matter field $\phi$ in a ``background'' metric\footnote{Recall, however, that the reference
metric $\hg_\mn$ that enters only the conformal parametrization of 2D metrics is to be distinguished carefully from the
true background metric $\gb_\mn$ which is at the heart of the entire gravitational EAA setting. In this conformal
parametrization, a generic bi-metric action $F[g,\gb]$ translates into a functional of \emph{two} conformal factors,
$F\big[\phi,\bar{\phi};\hg\big] \equiv F\big[\mku\hg\,\e^{2\phi},\mku\hg\,\e^{2\bar{\phi}}\mku\big]$.} $\hg_\mn$.
It would then be natural to ascribe to this field the stress tensor
\begin{equation}
 T_k^\text{L}[\phi;\hg]^\mn \equiv \frac{2}{\shg}\frac{\delta\GL[\phi;\hg]}{\delta\hg_\mn}\,.
\label{eq:TLiouvilleDef}
\end{equation}
Without using the equation of motion (i.e.\ ``off shell'') its trace is given by
\begin{equation}
 \Theta_k^\text{L}[\phi;\hg]\equiv \hg_\mn\,T_k^\text{L}[\phi;\hg]^\mn = \frac{\cgr}{12\pi}
  \left(\hB\phi + 2\mku\bls\,k^2\mku\e^{2\phi} \right).
\label{eq:TLiouville}
\end{equation}
Concerning \eqref{eq:TLiouville}, several points are to be noted.

\begin{enumerate}
\item Varying $\GL$ with respect to $\phi$ yields Liouville's equation $\hB\phi+2\mku\bls\, k^2\mku\e^{2\phi}
= \frac{1}{2}\hR$. With $\phi_\text{sol}$ denoting any solution to it we obtain ``on shell'' the following
$k$-independent trace:
\begin{equation}
 \Theta^\text{L}[\phi_\text{sol};\hg] = +\cgr \frac{1}{24\pi}\hR \,.
\label{eq:ThetaLiouville}
\end{equation}
If we now compare \eqref{eq:ThetaLiouville} to the general rule \eqref{eq:TCentralCharge} we conclude that the
Liouville field represents a CFT with the central charge
\begin{equation}
 c^\text{L} = -\cgr\,,
\end{equation}
which is \emph{negative} for pure asymptotically safe gravity, namely $c^\text{L} = -25$, or $-19$, respectively.
\end{enumerate}
Does this result indicate that the fixed point CFT is non-unitary, after all? The answer is a clear `no', and the
reason is as follows.
\begin{enumerate}[resume]
\item The Liouville theory governed by $\GL$ of \eqref{eq:LiouvilleNGFP} is not a faithful description of the
NGFP. According to eq.\ \eqref{eq:GammaGravILiou}, the full action contains the ``pure gravity'' term
$\frac{\cgr}{96\pi}I[\hg]$ in addition. In order to correctly identify the central charge of the NGFP it is essential
to add the $\hg_\mn$-derivative of this term to the Liouville stress tensor. Hence, the trace \eqref{eq:TLiouville}
gets augmented to
\begin{align}
 \frac{2\mku\hg_\mn}{\shg}\frac{\delta}{\delta\hg_\mn}\left(\frac{\cgr}{96\pi}I[\hg]\right)+\Theta_k^\text{L}[\phi;\hg]
 &= -\frac{\cgr}{24\pi}R(\hg)+\Theta_k^\text{L}[\phi;\hg] 
\label{eq:ThetaFull}\\
 &= \frac{\cgr}{24\pi}\left[-R(\hg)+2\mku\hB\phi+4\mku\bls\, k^2\mku\e^{2\phi}\right]
\nonumber\\
 &= \frac{\cgr}{24\pi}\left[-\e^{-2\phi}\big(R(\hg)-2\mku\hB\phi\big)+4\mku\bls\, k^2\right] \e^{2\phi}
\nonumber\\
 &= \frac{\cgr}{24\pi}\left[-R\big(\hg\mku\e^{2\phi}\big)+4\mku\bls\, k^2\right] \e^{2\phi}
\nonumber\\
 &= \e^{2\phi}\,\Theta_k\big[\mku\hg\mku\e^{2\phi}\big].
\nonumber
\end{align}
In the 2${}^\text{nd}$ line of \eqref{eq:ThetaFull} we inserted \eqref{eq:TLiouville}, in going from the 3${}^\text{rd}$
to the 4${}^\text{th}$ line we exploited the identity \eqref{eq:WeylR} from the appendix, and in the last line we used
\eqref{eq:ThetaParams}. So with this little calculation we have checked that the Liouville stress tensor makes physical
sense only when combined with the pure gravity piece.\footnote{In isolation, $\Theta^\text{L}[\phi;\hg]$ is not
invariant under the split-symmetry transformations \eqref{eq:SplitSymmetry}, i.e.\ not a function of the combination
$\hg\mku\e^{2\phi}$ only.} If this is done, the total gravitational trace from which the correct central charge is
inferred, eq.\ \eqref{eq:ThetaParams}, is indeed recovered, as it should be. It satisfies the relation\footnote{The
explicit factor $\e^{-2\phi}$ in \eqref{eq:ThetaAdded} is simply due to the different volume elements $\shg$ and
$\sg=\shg\,\e^{2\phi}$ appearing in the definitions of the stress tensors \eqref{eq:TLiouvilleDef} and
\eqref{eq:StressTensorGeneral}, respectively.}
\begin{equation}
 \Theta_k[\mku g] \equiv \Theta_k\big[\mku\hg\mku\e^{2\phi}\big] 
  = \e^{-2\phi}\bigg(-\frac{\cgr}{24\pi}\,\hR+\Theta_k^\text{L}[\phi;\hg]\bigg),
\label{eq:ThetaAdded}
\end{equation}
which holds true even off shell.

\item If we take $\phi$ on shell, eq.\ \eqref{eq:ThetaLiouville} applies and so the two terms in the brackets
of eq.\ \eqref{eq:ThetaAdded} cancel precisely. This, too, is as it should be since from eq.\ \eqref{eq:Rsc} we know
already that $\Theta_k[g]$ vanishes identically when $g\equiv\gb$ is a self-consistent background, and this is exactly
what we insert into \eqref{eq:ThetaAdded} when $\phi$ is a solution of Liouville's equation.

\end{enumerate}

Thus, taking the above points together we now understand that nothing is wrong with $c^\text{L} = -\cgr$. In fact,
$c^\text{L}<0$ for pure gravity is again a reflection of the Liouville field's ``wrong-sign'' kinetic
term\footnote{Hence, at the technical level, the wrong-sign kinetic term requires special attention (regularization,
analytic continuation, or similar) at intermediate steps of the calculation at most.} and its perfectly correct
property of reducing the total number of degrees of freedom.

\section{Status of different field parametrizations}
\label{sec:Emergence}

To fully establish the properties of the NGFP in exactly $2$ dimensions, it remains to understand the status and
reliability of the exponential and linear field parametrizations, respectively. Why is it the former that reproduces the
results of standard conformal field theory, including the bosonic critical dimension, while the latter fails to do so
\cite{N15,CD15,F15}?

\subsection{Different universality classes?}
\label{sec:UnivClasses}

It might be appropriate to begin with a word of warning: For the time being, it is not clear whether the exponential
and the linear parametrization, respectively, describe the same physics at the exact level. We cannot fully exclude the
possibility that both of them are equally correct, but probe instead two different universality classes.

It is a notoriously difficult question in virtually all functional integral based approaches\footnote{It is well known
that standard 1D configuration space functional integrals are dominated by non-differentiable paths since the set of
differentiable ones has measure $0$. The basic laws of quantum mechanics, noncommutativity of positions and momenta,
force us to include these classically forbidden non-differentiable trajectories in the path integral \cite{GCP15}.
Similarly, a consistent gravitational path integral might require integrating over ``metrics'' which have further
nonclassical features to a degree that is to be found out.} to quantum gravity
whether, or to what extent, \emph{degenerate, wrong-signature} or even \emph{vanishing} tensor fields should be
included \cite{W88}. Since the set of nondegenerate metrics with fixed signature forms a nonempty open subset in the
space of all covariant symmetric $2$-tensor fields \cite{DN15}, there is no a priori reason to expect that it has
vanishing measure, and so this question has no obvious answer.
It is known, however, that ``sufficiently different'' choices will lead to inequivalent theories \cite{AL98}.

In this context it might well be that a NGFP in a theory which is based only on nondegenerate fixed-signature metrics
falls into another universality class than a fixed point in a theory which includes all covariant symmetric tensors
without restrictions concerning degeneracy and signature. If so, the properties of the respective NGFP, in particular
the values of the central charge, could be different then.

Interestingly enough, the two parametrizations considered here give rise to these two different options, namely genuine
metrics vs.\ unconstrained symmetric tensors:

\noindent
\textbf{(a)}
In ref.\ \cite{DN15} it was argued that the exponential parametrization is adapted to the nonlinear structure of the
field space $\mF$ of all pure metrics, i.e.\ objects in $\mF$ are nondegenerate and have a prescribed signature. The
nondegeneracy and signature requirements prevent $\mF$ from exhibiting vector space character. As a consequence,
it is not possible to simply \emph{add} an arbitrary fluctuation to a given metric in order to obtain another metric
since it may happen that one leaves the set $\mF$ owing to the addition. To reach another metric one rather has to start
from a given metric and \emph{move along geodesics}. Only then it is ensured that all metrics parametrized this way
satisfy the signature and nondegeneracy constraint. It can be shown that there is a canonical connection on $\mF$
adjusted to this constraint, and that the resulting geodesics are parametrized precisely by the exponential relation
$g_\mn=\gb_{\mu\rho} (\e^h)^\rho{}_\nu$ with a symmetric tensor field $h_\mn$ \cite{DN15}. This explains the special
status of the exponential parametrization.

\noindent
\textbf{(b)}
The linear parametrization, $g_\mn=\gb_\mn+h_\mn$, on the other hand, allows integrating over a much larger space of
field configurations including degenerate, wrong-signature and vanishing tensor fields:\footnote{This would be in the
spirit of ref.\ \cite{W88}, and one might expect to find a phase of unbroken diffeomorphism invariance, among others.}
Since $h_\mn$ can be any symmetric tensor field, it is clear that the set of pure metrics can be left by adding it to
$\gb_\mn$.

Thus, the two parametrizations fall indeed into qualitatively different categories, and there is at least the
possibility that future RG studies might show that the respective NGFPs refer to different universality classes. We
conjecture that these classes would then be represented by $\cgr=19$ for the linear parametrization and by $\cgr=25$ if
the metric is parametrized by an exponential. In any case, the issue of parametrization dependence \cite{GKL15} should
always be reconsidered when a better truncation becomes technically manageable.\footnote{A first indication pointing
towards the possibility of different classes might be contained in recent results from the $f(R)$-truncation in 4D
where an apparently parametrization dependent number of relevant directions was observed \cite{OPV15}.}

We believe that although at this stage no parametrization of the metric should be preferred over the other one, the
compatibility of EAA-based Asymptotic Safety results with other approaches to quantum gravity can in principle
discriminate between different parametrizations. The critical value `25' for the central charge indicates indeed that
our calculations based on the exponential parametrization are closer to those of conformal field theory.
In the next subsection we aim at presenting another argument suggesting that the exponential parametrization is
particularly appropriate in the 2D limit.

\subsection{The birth of exponentials in 2D}
\label{sec:GoodVsBad}

The argument in this subsection considers only such dynamical metrics $g_\mn$ that are conformally related to a fixed
reference metric $\hg_\mn$, and only their relative conformal factor is quantized. The resulting ``conformally
reduced'' setting \cite{RW09,RW09b} amounts to the exact theory in 2D, but it is an approximation in higher dimensions.
(Accordingly, ``exponential parametrization'' refers to the form of the conformal factor in the following.)
Among all possible ways of parametrizing the conformal factor there exists a distinguished parametrization in each
dimension $d$.

\medskip
\noindent
\textbf{(1) Distinguished parametrizations.} Let us consider the conformal reduction of the Einstein--Hilbert action
$\SEH[g]\equiv -\frac{1}{16\pi G}\int\dd x\sg\,(R-2\Lambda)$ in any number of dimensions $d>2$. That is, we evaluate
$\SEH$ only on metrics which are conformal to a given $\hg$ consistent with the desired topology. But how should we
write the factor relating $g$ and $\hg$ now? Assume, for instance, the reduced $\SEH$ plays the role of a bare action
under a functional integral over a certain field $\Omega$ representing the conformal factor, how then should the latter
be written in terms of $\Omega\,$? Clearly, infinitely many parametrizations are possible here, and
depending on our choice the reduced $\SEH$ will look differently.

There exists a distinguished parametrization, however, which is specific to the dimensionality $d$, having the property
that $\int\!\sg\,R$ \emph{becomes quadratic in} $\Omega$. Starting out from a power ansatz, $g_\mn=\Omega^{2\nu}\,
\hg_\mn$, the integral $\int\!\sg\,R$ will in general produce a potential term $\propto \hR$ times a particular power
of $\Omega$, and a kinetic term $\propto \big( \hD\Omega \big)^2$ times another power of $\Omega$. The exponent of the
latter turns out to be zero, yielding a kinetic term quadratic in $\Omega$, precisely if \cite{JNP05}
\begin{equation}
 \nu = 2/(d-2)\,,\qquad g_\mn = \Omega^{4/(d-2)}\,\hg_\mn \,.
\label{eq:ExponentChoice}
\end{equation}
In this case the potential term is found to be quadratic as well, and one obtains \cite{JNP05,RW09}
\begin{equation}[b]
 \SEH[g = \Omega^{4/(d-2)}\,\hg] = -\frac{1}{8\pi G}\int\dd x\shg\left[ \frac{1}{2\,\xi(d)}\,\hD_\mu\Omega\,
  \hD^\mu \Omega + \frac{1}{2}\hR\,\Omega^2-\Lambda\,\Omega^{2d/(d-2)}\right].
\label{eq:SCREH}
\end{equation}
Here we introduced the constant
\begin{equation}
 \xi(d) \equiv \frac{(d-2)}{4(d-1)}\,.
\end{equation}
Usually one employs $\Omega(x)-1\equiv\omega(x)$ rather than $\Omega$ itself as the dynamical field that is
quantized, i.e.\ integrated over if $\SEH$ appears in a functional integral. Then there will be no positivity issues
as long as $\omega(x)$ stays small. We emphasize, however, that the derivation of neither \eqref{eq:SCREH} nor the
related action for $\omega$,
\begin{equation}
 \SEH[\omega;\hg] = -\frac{1}{8\pi G}\int\dd x\shg\left[ \frac{1}{2\,\xi(d)}\,\hg^\mn \p_\mu\omega\,\p_\nu \omega
  + \frac{1}{2}\hR\,(1+\omega)^2-\Lambda\,(1+\omega)^{2d/(d-2)}\right],
\label{eq:omegaAction}
\end{equation}
involves any (small field, or other) expansion. (It involves an integration by parts though, hence there could be
additional surface contributions if spacetime has a boundary.)

\medskip
\noindent
\textbf{(2) Metric operators.} The exponent appearing in the conformal factor $\Omega^{2\nu}$ is
non-integer in general, exceptions being $d=3,4$, and $6$, see Table \ref{tab:Dimensions}.
{\renewcommand{\arraystretch}{1.2}
\begin{table}[tp]
\centering
\begin{tabular}{cccc}
 \hline
 $d$  &  $\quad 3\quad$  &  $\quad 4\quad$  &  $\quad 6\quad$\\
 \hline
 $\;$ Conformal factor $\;$  &  $\Omega^4$  &  $\Omega^2$  &  $\Omega$\\
 Volume operator &  $\Omega^6$  &  $\Omega^4$  &  $\Omega^3$\\
 \hline
\end{tabular}
\caption{Conformal factor and volume operator for the distinguished parametrization.}
\label{tab:Dimensions}
\end{table}%
}%
The virtue of a quadratic action needs no mentioning, of course. As long as the cosmological constant plays no
role --- $\Lambda$ will always give rise to an interaction term --- the computation of the RG flow will be easiest and
\emph{most reliable} if we employ the distinguished parametrization.\footnote{The RG flow of
the conformally reduced Einstein--Hilbert truncation (``CREH'') with the distinguished parametrizations has been
computed in \cite{RW09}, an LPA-type extension was considered in \cite{RW09b}, see also \cite{MP09}.}

One should be aware that there is a conservation of difficulties also here. Generically the conformal factor depends on
the quantum field \emph{nonlinearly}. Hence, canonically speaking, even if the action is trivial (Gaussian), the
construction of a \emph{metric operator} amounts to defining $\Omega^{2\nu}$ or $(1+\omega)^{2\nu}$ as a composite
operator. And in fact, the experience with models such as Liouville theory \cite{OW86,DO94,KN93} shows how extremely
difficult this can be.

At present we are just interested in comparing the relative degree of reliability of two truncated RG flows, based upon
different field parametrizations. For this purpose it is sufficient to learn from the above argument that the ``most
correct'' results should be those from the distinguished parametrization \eqref{eq:ExponentChoice} since then the
theory is free (for $\Lambda=0$). But what is the distinguished parametrization in $2$ dimensions?

\medskip
\noindent
\textbf{(3) The limit \bm{$d\rightarrow 2$}}. As we lower $d$ towards two dimensions, the distinguished form of the
conformal factor, $(1+\omega)^{4/(d-2)}$, develops into a function which increases with $\omega$ faster than any
power. At the same time the constant $\xi(d)$ goes to zero, and \eqref{eq:omegaAction} becomes
\begin{equation}
\begin{split}
 \SEH[\omega;\hg] = -\frac{1}{16\pi \bG}\int\td^{2+\ve}x\shg\,\bigg[\frac{4}{\ve^2}\,\hg^\mn \p_\mu\omega\,
  \p_\nu\omega\,\big\{1+\mO(\ve)\big\} & \\
  + \frac{1}{\ve}\,\hR\,(1+\omega)^2 - 2\bL\,(1+\omega)^{2(2+\ve)/\ve}& \bigg].
\end{split}
\label{eq:omegaAction2}
\end{equation}
Here we wrote $G\equiv \bG\,\ve$ and $\Lambda\equiv \bL\,\ve$ again, assuming that $\bG,\bL=\mO(\ve^0)$. We see that in
order to obtain a meaningful kinetic term we must rescale $\omega$ by a factor of $\ve$ prior to taking the limit
$\ve\searrow 0$.

Introducing the new field $\phi(x) \equiv 2\omega(x)/\ve$, its kinetic term $\hg^\mn \p_\mu\phi\,
\p_\nu\phi\,\big\{1+\mO(\ve)\big\}$ will have a finite and nontrivial limit. The concomitant conformal factor
$\Omega^{2\nu}$ has the limit
\begin{equation}
 \lim\limits_{\ve\rightarrow 0} \, (1+\omega)^{4/\ve}
 = \lim\limits_{\ve\rightarrow 0}  \Big(1+{\textstyle\frac{1}{2}}\ve\phi\Big)^{4/\ve}
 = \lim\limits_{n\rightarrow\infty}  \Big(1+{\textstyle\frac{2\phi}{n}}\Big)^n = \e^{2\phi} \,.
\end{equation}
This demonstrates that \emph{the exponential parametrization} $g_\mn = \e^{2\phi}\hg_\mn$ \emph{is precisely the 2D
limit of the distinguished (power-like) parametrizations in} $d>2$.

The cosmological term in \eqref{eq:omegaAction2} involves the same exponential for $d\rightarrow 2$, and the originally quadratic potential
$\hR(1+\omega)^2$ turns into a linear one for $\phi$. Taking everything together the Laurent series of $\SEH$ in $\ve$
looks as follows:
\begin{equation}
 \SEH[\phi;\hg] = -\frac{1}{16\pi\bG}\bigg\{ \frac{1}{\ve}\int\!\td^{2+\ve} x\shg\,\hR + \int\!\td^2 x\shg\,
 \Big(\hg^\mn\,\p_\mu\phi\,\p_\nu\phi + \hR\,\phi - 2\bL\,\e^{2\phi}\Big)\bigg\} + \mO(\ve).
\label{eq:SEHphiAction}
\end{equation}
The first term on the RHS is $\phi$-independent and involves a purely topological contribution proportional to the
Euler characteristic, which was already encountered in section \ref{sec:IndGravityFromEH}. Obviously from
\eqref{eq:SEHphiAction} we reobtain Liouville theory as the intrinsically 2D part of the Einstein--Hilbert action, but
this is perhaps not too much of a surprise.

What is important, though, is that in this derivation, contrary to the discussion in the previous sections, the
exponential field dependence of the conformal factor was not put in by hand, we rather \emph{derived} it.

Here our input
were the following two requirements: First, the scaling limit of $\SEH$ should be both non-singular and nontrivial, and
second, it should go through a sequence of actions which, apart from the cosmological term, are at most quadratic in the
dynamical field. Being quadratic implies that when $\SEH[\omega;\hg]$ is used as the (conformal reduction of the)
Einstein--Hilbert truncation, this truncation is ``perfect'' at any $\ve$.

Therefore, we believe that using the exponential parametrization already in ``higher'' dimensions $d>2$ yields
more reliable results for the $\beta$-functions and their 2D limits than using the linear parametrization in $d>2$ and
taking the 2D limit of the corresponding $\beta$-functions.
(There is still a minor source of uncertainty due to the ghost sector. In either parametrization there are
ghost-antighost-graviton interactions which are not treated exactly by the truncations we use.)

The basic difference between the two parametrizations can also be seen quite directly. If we insert $g=\e^{2\phi}\hg$
into $\SEH$, the resulting derivative term reads exactly, i.e.\ without any expansion in $\ve$ and/or $\phi$ and
rescaling of $\phi$:
\begin{equation}
 -\frac{(d-1)}{16\pi\bG}\int\dd x\shg\;\e^{(d-2)\phi}\big(\hD\phi\big)^2 \,.
\end{equation}
For $d\rightarrow 2$ this term has a smooth limit (we did use $G=\bG\,\ve$ after all) and this limit is quadratic in
$\phi$.

On the other hand, inserting the linear parametrization $g=(1+\omega)\hg$ into $\SEH$ we obtain again exactly, i.e.\
without expanding in $\ve$ and/or $\omega$ and rescaling $\omega$:
\begin{equation}
 -\frac{(d-1)}{64\pi\bG}\int\dd x\shg\;(1+\omega)^{(d-2)/2}\,\frac{\big(\hD\omega\big)^2}{(1+\omega)^2} \;.
\label{eq:KinTermOmega}
\end{equation}
The term \eqref{eq:KinTermOmega}, too, has a smooth limit $d\rightarrow 2$, but it is not quadratic in the dynamical
field. This renders the $\omega$-theory interacting and makes it a nontrivial challenge for the truncation.

\medskip
\noindent
\textbf{(4) The dimension \bm{$d=6$}}. As an aside we mention that according to Table \ref{tab:Dimensions} the case
$d=6$ seems to be easiest to deal with since in the preferred field parametrization the conformal factor is linear in
the quantum field, and so there is no need to construct a composite operator. The kinetic term \eqref{eq:KinTermOmega}
becomes quadratic exactly at $d=6$.

It is intriguing to speculate that this observation is related to the following rather surprising property enjoyed by
the $\beta$-functions derived from the bi-metric Einstein Hilbert truncation (see Appendix A.1 of ref.\ \cite{BR14}):
If $d=6$, and if in addition the dimensionful dynamical cosmological constant $\Lambda^\text{Dyn}$ is zero, then
\emph{the gravity contributions to the $\beta$-functions of both $\Lambda^\text{Dyn}$ and the dimensionful dynamical
Newton constant $G^\text{Dyn}$ vanish exactly}. (There are nonzero ghost contributions, though.)

\medskip
\noindent
\textbf{(5) Summary}. On the basis of the above arguments we conclude that \emph{most probably the exponential
pa\-ra\-me\-tri\-za\-tion is more reliable in 2D than the linear one}. We believe in particular that $\cgr=25$ is more
likely to be a correct value of the central charge at the pure gravity fixed point than its competitor `19'. Depending
on the reliability of the linear parametrization, the `19' could be a poor approximation to `25', or a hint at another
universality class.

\section{The reconstructed functional integral}
\label{sec:recFI}

For the following it is important to keep in mind that the ``derivation'' of the FRGE from a functional integral is
only formal as it ignores all difficulties specific to the UV limit of quantum field theories. In fact, rather than the
integral, the starting point of the EAA based route to a fundamental theory is the mathematically perfectly well
defined, UV cutoff-free flow equation \eqref{eq:FRGE}. In this setting, the problem of the UV limit is shifted from the
properties of the equation itself to those of its \emph{solutions}, converting renormalizability into a condition on
the existence of fully extended RG trajectories on theory space. The Asymptotic Safety paradigm is a way of achieving
full extendability in the UV and, barring other types of (infrared, etc.) difficulties, it leads to a well-behaved
action functional $\Gamma_k$ at each $k\in[0,\infty)$.

\subsection{The reconstruction process}

While every complete RG trajectory defines a quantum field theory (with the cutoff(s) removed), it does \emph{not} give
rise to a functional integral description of this theory a priori. Nevertheless, one may ask for a functional integral
reproducing a given complete $\Gamma_k$-trajectory. This ``reconstruction'' step has been motivated and discussed in
detail in ref.\ \cite{MR09}.

The association of a functional integral, i.e.\ a bare theory, to a $\Gamma_k$-trajectory is highly non-unique. The
first decision to be taken concerns the variables of integration: They may or may not be fields of the same sort as
those serving as arguments of $\Gamma_k$. From the practical point of view the most important situation is when the
integration variables are no (discretized) fields at all, but rather belong to a certain statistical mechanics model
whose partition function at criticality is supposed to reproduce the predictions of the EAA trajectory. Besides the
nature of the integration variables, a UV regularization scheme, a correspondingly regularized functional integration
measure, and an associated bare action $\SB$ are to be chosen. Then the information encapsulated in
$\Gamma_{k\rightarrow\infty}$ can be used in order to find out how the bare parameters contained in $\SB$ must depend
on the UV cutoff $\UV$ in order to give rise to a well-defined path integral reproducing the EAA-trajectory in the
limit $\UV\rightarrow\infty$.

In \cite{MR09} this reconstruction has been carried out explicitly for metric gravity in the Einstein--Hilbert
truncation for the following choices:
\textbf{(i)}
The integration variable is taken to be the metric fluctuation $h_\mn$, i.e.\ (accidentally) the same sort of
variable as in the argument of $\Gamma_k$.
\textbf{(ii)}
The UV regularization is implemented by means of a sharp mode cutoff.\footnote{When the cosmological constant is
an issue a higher-derivative regularization in the UV similar to the one in the IR is known to be problematic
\cite{MR09}. For a purely scalar theory without gauge and gravity fields the implications of a sharp cutoff were
discussed in ref.\ \cite{MS15} recently.}
\textbf{(iii)}
The truncated bare action $\SB$ is of Einstein--Hilbert type with bare couplings $\cg_\UV$ and $\cl_\UV$,
respectively. Under these conditions an explicit map relating $(g_k,\lambda_k)$ for $k\rightarrow\infty$ to
$(\cg_\UV,\cl_\UV)$ when $\UV\rightarrow\infty$ was derived.

This map depends on a parameter $M$ which labels a certain 1-parameter family of measures. This $M$-dependence reflects
the fact that it is not the bare action alone which is uniquely determined but rather the combination of measure and
bare action: Certain redefinitions of the measure can be absorbed by redefinitions of the bare action and vice versa,
signaling the ``unphysicalness'' of the bare action.

The $M$-dependence can be exploited to conveniently adjust the bare coupling constants. In $d=2+\ve$ dimensions there
is one particular value of $M$ that leads to an exactly vanishing bare cosmological constant,
$\cl_*\equiv \cl_{\UV\rightarrow\infty}=0$, and a bare Newton constant $\cg_\UV$ which equals precisely the effective
one at the NGFP \cite{NR16}:
\begin{equation}
 \cg_*=g_*\,.
\label{eq:BareEqualsEffective}
\end{equation}
After having reconstructed the gravitational functional integral in $d=2+\ve$, we take its 2D limit employing the
method of section \ref{sec:IndGravityFromEH}. In combination with eq.\ \eqref{eq:BareEqualsEffective} we obtain
\begin{equation}[b]
 \SB^\text{grav}[g] = \frac{(25-\ns)}{96\pi} \,I[g] +\cdots
\label{eq:SBareGrav}
\end{equation}
The dots indicate that there might appear additional terms originating from the zero modes, according to eq.\
\eqref{eq:LimitResultGen} in the appendix.
Further details about the reconstruction and the derivation of \eqref{eq:BareEqualsEffective} will be
presented elsewhere \cite{NR16}. For our present purposes it suffices to know that the bare action has the structure
\eqref{eq:SBareGrav}, displaying a cutoff independent term $\propto I[g]$ and possibly ``zero mode terms''.

The bare action of the matter system too can be reconstructed according to the results of ref.\ \cite{MS15}: For
cutoffs satisfying certain constraints the bare action equals precisely the EAA when the respective cutoff scales are
identified. Thus, the bare matter action is given by the RHS of eq.\ \eqref{eq:matter}.

We would like to point out that, as a consequence, the number $\ns$ enters \emph{both the matter and the gravitational
part of the bare action}. (Note that the fixed point value $g_*$ depends on $\ns$, cf.\ sec.\ \ref{sec:LimitFullEH}.)

\subsection{A functional integral for 2D asymptotically safe gravity}

\textbf{(1) The partition function}. Based on the above considerations we obtain the full reconstructed partition
function
\begin{equation}
 Z = \int[\td\tau] \int\mD_{\gfull}\phi\;\, Z_\text{gh}\big[\gfull\big]\,\Zm\big[\gfull\big]\, \Yg\big[\gfull\big]\,.
\label{eq:PartFunc}
\end{equation}
The integrand of \eqref{eq:PartFunc} comprises the following factors:
the exponential of the gravitational part of the fixed point action,
\begin{equation}
 \Yg[g] \equiv \exp\!\left(-\frac{(25-\ns)}{96\pi}\,I[g]+\cdots\right),
\label{eq:Ygrav}
\end{equation}
the partition function of the matter system,
\begin{equation}
\begin{split}
 \Zm[g] &\equiv \int\mD A\;\exp\!\Bigg(-\frac{1}{2}\sum\limits_{i=1}^{\ns} \int \td^2 x\sg\,
  g^\mn\mku\p_\mu A^i\,\p_\nu A^i\Bigg)\\
	&= {\det}^{-\ns/2}\big(-\Box_g\big)
	= \exp\left(-\frac{\ns}{96\pi}\,I[g]+\cdots\right),
\end{split}
\label{eq:Zmatter}
\end{equation}
the partition function of the $b$-$c$ ghost system, $Z_\text{gh}$, the split symmetry invariant measure for the
integration over the Liouville field, $\mD_{\gfull}\phi$, and finally the measure $[\td\tau]$ for the integration over
the moduli that are implicit in the reference metric pertaining to a given topological type of the spacetime manifold
(cf.\ sec.\ \ref{sec:ConfGauge}). In eqs.\ \eqref{eq:Ygrav} and \eqref{eq:Zmatter} we suppressed possible contributions
to the bare cosmological constant. Here and in the following we indicate them by the dots.

The behavior under Weyl transformations of the various factors is well known. Using in particular
eq.\ \eqref{eq:decomp1} with the (non-cosmological constant part of the) renormalized Liouville action, $\Delta I$, as
defined in \eqref{eq:DeltaIDef}, we have
{\allowdisplaybreaks
\begin{subequations}
\begin{align}
 \Yg\big[\gfull\big] &= \Yg[\mku\hg]\;\exp\!\bigg(\!+\frac{(25-\ns)}{12\pi}\,\Delta I[\phi;\hg]\bigg), \\
 \Zm\big[\gfull\big] &= \Zm[\mku\hg]\;\exp\!\bigg(\!+\frac{\ns}{12\pi}\,\Delta I[\phi;\hg]\bigg), \\
 Z_\text{gh}\big[\gfull\big] &= Z_\text{gh}[\mku\hg]\;\exp\!\bigg(\!+\frac{(-26)}{12\pi}\,\Delta I[\phi;\hg]\bigg),
 \label{eq:WeylGhosts}\\
 \mD_{\gfull}\phi &= \mD_{\hg}\phi\;\exp\!\bigg(\!+\frac{1}{12\pi}\,\Delta I[\phi;\hg]\bigg).
 \label{eq:WeylDPhiMeasure}
\end{align}
\label{eq:WeylPartition}
\end{subequations}
}%
As before, possible (measure dependent) terms involving the bare cosmological constant are suppressed in eqs.\
\eqref{eq:WeylPartition}. On the RHS of \eqref{eq:WeylDPhiMeasure}, $\mD_{\hg}\phi$ is the translational invariant
measure now.

Up to this point, the discussion is almost the same as in non-critical string theory \cite{P81}. The profound
difference lies in the purely gravitational part of the bare action, $\Yg$. Contrary to what happens in any
conventional field theory, whose bare action is a \emph{postulate} rather than the result of a \emph{calculation},
asymptotically safe gravity in $2$ dimensions is based upon a gravitational action \emph{which depends explicitly on
properties of the matter system}. In the example at hand, this dependence is via the number $\ns$ of $A^i$-fields that
makes its appearance in the fixed point action and hence in the ``Boltzmann factor'' \eqref{eq:Ygrav}.

\medskip
\noindent
\textbf{(1a) Matter refuses to matter: a compensation mechanism}. Remarkably enough, the integrand of
\eqref{eq:PartFunc} depends on $\ns$ only via the product $\Zm\cdot\Yg$ in which the $\ns$-dependence cancels between
the two factors. Multiplying \eqref{eq:Ygrav} and \eqref{eq:Zmatter} we obtain a result which, for any $\ns$, equals
that of pure gravity. It is always the same no matter how many scalar fields there are:
\begin{equation}
 \Zm[g]\,\Yg[g]= \exp\!\left(-\frac{25}{96\pi}\,I[g]+\cdots\right),
\end{equation}
which transforms as $\Zm\big[\gfull\big]\mku\Yg\big[\gfull\big]=\Zm[\hg]\mku\Yg[\hg]\,
\exp\left(+\frac{25}{12\pi}\Delta I[\phi;\hg]\right)$ under a Weyl rescaling. As a consequence, the reconstructed
functional integral coincides always with that of \emph{pure gravity} (as long as we do not evaluate the expectation
value of observables depending on the $A$'s and as long as cosmological constant terms do not play a role):
\begin{equation}[b]
 \!Z = \int[\td\tau]\;\Zm[\hg]\,\Yg[\hg] \int\!\mD_{\gfull}\phi\;\, Z_\text{gh}\big[\gfull\big]\,
 \exp\left(\!+\frac{25}{12\pi}\,\Delta I[\phi;\hg]+\cdots\mkern-1mu\right).\!\!
\end{equation}

\medskip
\noindent
\textbf{(1b) Zero total central charge}. Over and above the specific form of its matter dependence, the fixed point
action displays a second miracle: Its central charge equals precisely the critical value $25$. Up to a cosmological
constant term possibly, this leads to a complete cancellation of the entire $\phi$-dependence of the integrand once the
ghost contribution \eqref{eq:WeylGhosts} and the ``Jacobian'' factor in \eqref{eq:WeylDPhiMeasure} are taken into
account:
\begin{equation}[b]
 Z = \int[\td\tau]\;Z_\text{gh}[\hg]\,\Zm[\hg]\,\Yg[\hg] \int\mD_{\hg}\phi \; \exp(0+\cdots)\,.
\label{eq:ZNonCritical}
\end{equation}
Hence, for every choice of the matter sector, the total system described by the reconstructed functional integral of
asymptotically safe 2D gravity is a conformal field theory with central charge zero. The various sectors of this system
contribute to the total central charge as follows:
\begin{equation}
 c_\text{tot} = \underbrace{(25-\ns)}_{\text{NGFP, grav.\ part}}+\underbrace{\phantom{|}\ns\phantom{|}}_{\text{matter}}
 + \underbrace{\phantom{|}1\phantom{|}}_{\text{Jacobian}} + \underbrace{(-26)}_{\text{ghosts}} = 0 \,.
\label{eq:ctotZero}
\end{equation}

Actually the result \eqref{eq:ctotZero} is even more general than we indicated so far. In addition to the scalar matter
fields underlying our considerations up to this point we can also bring massless free Dirac fermions into play and
couple them (minimally) to the dynamical metric by adding a corresponding term to the matter action \eqref{eq:matter}.
The contribution of each of such fermions to the $\beta$-function of Newton's constant in $d=2+\ve$ is the same as for
a scalar field \cite{DP13}, that is, fermions and scalars enter the central charge in the same way. Hence, in all above
equations for $\beta$-functions and central charges we may identify $\ns$ with
\begin{equation}
 \ns\equiv N_\text{S}+N_\text{F}\,,
\end{equation}
where $N_\text{S}$ and $N_\text{F}$ denote the number of real scalars and Dirac fermions, respectively.
In particular, we recover the same cancellation in the total central charge as in eq.\ \eqref{eq:ctotZero}: The central
charge of the matter system, $+\ns$, removes exactly a corresponding piece in the pure gravity contribution enforced by
the fixed point, $25-\ns$.

\medskip
\noindent
\textbf{(2) Observables}. By inserting appropriate functions $\bar{\ob}[\phi,A;\hg]$ into the path integral
\eqref{eq:PartFunc} we can in principle evaluate the expectation values of arbitrary observables $\ob[\mku g,A]=
\ob[\gfull,A]$. In the case when the observables do not involve the matter fields, their expectation
values read
\begin{equation}
 \langle\ob\rangle=\frac{1}{Z}\int[\td\tau]\;\Zm[\hg]\,\Yg[\hg] \int\mD_{\gfull}\phi\;\, Z_\text{gh}\big[\gfull\big]\,
 \bob[\phi;\hg]\,\exp\left(\frac{25}{12\pi}\,\Delta I[\phi;\hg]\right).
\label{eq:expObs}
\end{equation}
Without actually evaluating the $\phi$-integral we see that when the cosmological constant term is negligible \emph{the
expectation value of purely gravitational observables does not depend on the presence or absence of matter and its
properties}, provided the background factor $\Zm[\hg]$ in \eqref{eq:expObs} cancels against the corresponding piece in
the denominator of \eqref{eq:expObs}. At the very least, this happens if one considers expectation values at a fixed
point of moduli space.

\medskip
\noindent
\textbf{(3) Gravitational dressing}. As it is well known \cite{DDK88,Wa93}, it is not completely straightforward to
find the functional $\bob[\phi;\hg]$ which one must use under a conformally gauge-fixed path integral in order to
represent a given diffeomorphism (and, trivially, Weyl) invariant observable $\ob[g]=\ob[\gfull]$. The association
$\ob\rightarrow\bob$ should respect the following conditions \cite{Wa93}: $\bob[\phi;\hg]$ must be invariant under
diffeomorphisms, it must approach $\ob[\gfull]$ in the classical limit and $\ob[\hg]$ in the limit $\phi\rightarrow 0$,
and most importantly it must be such that the expectation value computed with its help is independent of the reference
metric chosen, $\hg_\mn$.

Let us briefly recall the David--Distler--Kawai (DDK) solution to this problem \cite{DDK88}. For this purpose we
consider 2D gravity coupled to an arbitrary matter system described by a CFT with central charge $c$ and partition
function $\ZMc[g]$. First we want to evaluate the partition function for a fixed volume (area) of spacetime, $V$:
\begin{equation}
 Z_V=\int\frac{\mD g}{\text{vol(Diff)}}\;\ZMc[g]\;\delta\!\left(V-\int\td^2 x\sg\right).
\end{equation}
This integral involves the observable $\ob[g]\equiv\int\td^2 x\sg\equiv\int\td^2 x\shg\,\exp(2\phi)$. The associated
$\bob$ satisfying the above conditions turns out to require only a ``deformation'' of the prefactor of $\phi$ in the
exponential:
\begin{equation}
 \bob[\phi;\hg] = \int\td^2 x\shg\,\exp(2\mku\alpha_1 \phi)\,.
\label{eq:AreaOp}
\end{equation}
The modified prefactor $\alpha_1$ depends on the central charge of the matter CFT according to
\begin{equation}
 \alpha_1 = \frac{2\mku\sqrt{25-c}}{\sqrt{25-c}+\sqrt{1-c}} = \frac{1}{12}\left[25-c-\sqrt{(25-c)(1-c)}\right]\,.
\label{eq:AlphaOne}
\end{equation}
Thus, in the conformal gauge, $Z_V$ reads as follows:
\begin{equation}
 Z_V = \int[\td\tau]\;Z_\text{gh}[\hg]\,\ZMc[\hg] \int\mD_{\hg}\phi\;\delta\!\left(
  V-\int\td^2 x\shg\,\e^{2\mku\alpha_1\phi}\right)\;\exp\!\left(\!-\frac{(25-c)}{12\pi}\,\Delta I[\phi;\hg]\right).
\label{eq:PartFuncZV}
\end{equation}
Similarly the expectation value of an arbitrary observable $\ob[g]$ at fixed volume is given by $\langle\ob[g]\rangle
= Z_V^{-1}\langle\bob[\phi;\hg]\rangle'$. Here $\langle\cdots\rangle'$ is defined by analogy with \eqref{eq:PartFuncZV}
but with the additional factor $\bob[\phi;\hg]$ under the $\phi$-integral.

The DDK approach to the gravitational dressing of operators from the matter sector was developed as a conformal
gauge-analogue to the work of Knizhnik, Polyakov and Zamolodchikov (KPZ) \cite{KPZ88} based upon the light cone gauge.

To study gravitational dressing, let us consider an arbitrary spinless primary field $\ob_n[g] \equiv \int\td^2 x\sg\;
\mP_{n+1}(g)$, where $\mP_n(g)$ is a generic scalar involving the matter fields with conformal weight $(n,n)$, that is,
it responds to a rescaling of the metric according to $\mP_n(\e^{-2\sigma}g) = \e^{2 n \sigma}\,\mP_n(g)$. Under the
functional integral, the observables $\ob_n$ are then represented by
\begin{equation}
 \bob_n[\phi;\hg] = \int\td^2 x\shg\;\exp(2\mku\alpha_{-n}\mku\phi)\,\mP_{n+1}(\hg)\,,
\label{eq:OpN}
\end{equation}
where the $c$-dependent constants in the dressing factors generalize eq.\ \eqref{eq:AlphaOne}:
\begin{equation}
 \alpha_n = \frac{2\mku n\mku\sqrt{25-c}}{\sqrt{25-c}+\sqrt{25-c-24n}}
\label{eq:AlphaN}
\end{equation}
Using \eqref{eq:AlphaN} it is straightforward now to write down the modified conformal dimensions corrected by the
quantum gravity effects.

The results of the DDK approach reproduce those of KPZ (valid for spherical topology) and generalize them for
spacetimes of arbitrary topology. Within the framework of the EAA and its functional RG equations, the KPZ relations
were derived from Liouville theory in ref.\ \cite{RW97}; for a review see \cite{CD15}.

\medskip
\noindent
\textbf{(4) Quenching of the KPZ scaling}. Let us apply the general DDK--KPZ formulae to the NGFP theory of
asymptotically safe gravity. We must replace then
\begin{equation}
 c\;\longrightarrow\; \cgr + N \equiv (25-N)+N = 25\,,
\end{equation}
since the relevant bare action arises now from both the integrated-out matter fluctuations and the pure-gravity NGFP
contribution, $\Yg$. Setting $c=25$ in eqs.\ \eqref{eq:AlphaOne} and \eqref{eq:AlphaN} we obtain
\begin{equation}
 \alpha_1=0\qquad \text{and}\qquad \alpha_n=0 \mku ,
\label{eq:AlphaZero}
\end{equation}
respectively. This implies that \emph{the Liouville field completely decouples from the area operator \eqref{eq:AreaOp}
and any of the observables \eqref{eq:OpN}}.

As a consequence, the dynamics of the matter system is unaffected by its coupling to quantum gravity. In particular,
its critical behavior is described by the properties (critical exponents, etc.) of the matter CFT defined on a
non-dynamical, rigid background spacetime. Thus, the specific properties of the NGFP lead to a perfect ``quenching''
of the a priori expected KPZ scaling.

\medskip
\noindent
\textbf{(5) Relation to non-critical string theory}. The functional integral \eqref{eq:ZNonCritical} is identical to
the partition function of non-critical string theory in 25 Euclidean dimensions. This theory is equivalent to the usual
critical bosonic string living in a (25+1)-dimensional Minkowski space whereby the Liouville mode plays the role of
the time coordinate in the target space \cite{DG87,DNW89}. Whether we consider pure asymptotically safe gravity in two
dimensions, or couple any number of scalar and fermionic matter fields to it, the resulting partition function equals
always the one induced by the fluctuations of precisely 25 string positions $X^m(x^\mu)$.

There is, however, a certain difference between asymptotically safe gravity and non-critical string theory in the way
the special case of vanishing total central charge, i.e.\ of precisely 25 target space dimensions, is approached.
To see this note that in the present paper we related the Liouville field to the metric by eq.\ \eqref{eq:DefConfFac},
and at no point we redefined $\phi$ by absorbing any constant factors in it. In this connection the Liouville action
for a general central charge $c$ has the structure $\GL = -\frac{c}{24\pi} \int \big( \hD_\mu \phi \mku \hD^\mu \phi
+ \hR\phi \big) + \cdots$.

\noindent
\textbf{(i)} In order to combine $\GL$ with the action of the string positions, $+\frac{1}{8\pi}\int \hD_\mu X^m
\mku \hD^\mu X^m$, it is natural to introduce the redefined field
\begin{equation}
 \phi' \equiv Q\mku\phi\qquad \text{with}\qquad Q\equiv \sqrt{\frac{c}{3}} \,,
\end{equation}
in terms of which $\GL = -\frac{1}{8\pi}\int \big( \hD_\mu \phi' \mku \hD^\mu \phi' + Q\hR\phi' \big) + \cdots$.
It is this new field $\phi'$ that plays the role of time in target space and combines with the $X^m$'s in the
conventionally normalized action $\frac{1}{8\pi}\int \big(-\hD_\mu \phi' \mku \hD^\mu \phi' + \hD_\mu X^m
\mku \hD^\mu X^m - Q\hR\phi' \big) + \cdots$ which enhances the original $\mathrm{O}(25)$ symmetry to the full Lorentz
group in target space, $\mathrm{O}(1,25)$ \cite{DNW89}.

In string theory conformal invariance requires the total central charge to vanish, $c_\text{tot}=0$. Hence, arguing
that the combined $(X^0\equiv\phi',X^m)$-quantum system is equivalent to the usual bosonic string theory in the
critical dimension involves taking the limit $c\equiv c_\text{tot} \rightarrow 0$ in the above formulae. Obviously this
requires some care in calculating correlation functions as the relationship $\phi'\equiv \sqrt{c/3}\,\phi$ breaks down
in this limit. Considering vertex operators for the emission of a tachyon of 26-dimensional momentum $(P_0,P_m)$, say,
this involves combining the rescaling $\phi\rightarrow\sqrt{c/3}\,\phi$ with a corresponding rescaling of $P_0$ with
the inverse factor, $P_0\rightarrow\sqrt{3/c}\,P_0$, rendering their product $P_0 X^0\equiv P_0\phi'$ independent of
$c$. The vertex operator $\exp\mkern-1mu\big\{i(-P_0 X^0+P_m X^m)\big\}$, too, displays the full $\mathrm{O}(1,25)$
invariance. (See \cite{DG87} for a detailed discussion.)

\noindent
\textbf{(ii)} In 2D asymptotically safe quantum gravity, too, the total central charge was found to vanish, albeit for
entirely different reasons than in string theory. However, here there is no obvious reason or motivation for any
rescaling before letting $c\rightarrow 0$. In all of the above equations, including \eqref{eq:AreaOp} and
\eqref{eq:OpN}, $\phi$ still denotes the Liouville field introduced originally. In quantum gravity we let
$c\rightarrow 0$ in the most straightforward way, setting in particular $c=0$ directly in \eqref{eq:AlphaOne} and
\eqref{eq:AlphaN}. This is what led us to \eqref{eq:AlphaZero}, that is, the disappearance of $\phi$ from the
exponentials $\exp(2\mku\alpha_{-n}\phi)$ multiplying the matter operators and the ``quenching'' of the KPZ-scaling.

\subsection{Comparison with Monte Carlo results}

In earlier work \cite{DR09,LR05,RS11} indications were found that suggest that Quantum
Einstein Gravity in the continuum formulation based upon the EAA might be related to the discrete approach employing
Causal Dynamical Triangulation \cite{AL98,AGJL12}. In particular, the respective predictions for the fractal dimensions
of spacetime were compared in detail and turned out similar \cite{LR05,RS11}. It is therefore natural to
ask whether the quenching of the KPZ-scaling due to the above compensation mechanism can be seen in 2D CDT simulations.
And in fact, the Monte-Carlo studies indeed seem to suggest a picture that looks quite similar at first sight:
Coupling several copies of the Ising model \cite{AAL00} or the Potts model \cite{AALP09} to $2$-dimensional
Lorentzian quantum gravity in the CDT framework, there is strong numerical evidence that the critical behavior of the
combined system, in the matter sector, is described by \emph{the same} critical exponents as on a fixed, regular
lattice. Under the influence of the quantum fluctuations in the geometry the critical exponents do not get shifted
to their KPZ values.

While this seems a striking confirmation of our Asymptotic Safety-based prediction, one should be careful in
interpreting these results. In particular, it is unclear whether the underlying physics is the same in both cases.
In CDT, the presence (absence) of quantum gravity corrections of the matter exponents is attributed to the presence
(absence) of baby universes in Euclidean (causal Lorentzian) dynamical triangulations. In our approach instead, the
quantum gravity corrections that could in principle lead to the KPZ exponents are exactly compensated by the
\emph{explicit matter dependence} of the \emph{pure gravity}-part in the bare action. This matter dependence is an
immediate consequence of the very Asymptotic Safety requirement.

As yet we considered conformal matter only which was exemplified by massless, minimally coupled scalar fields. In the
non-conformal case when those fields are given a mass for instance, the compensation between the matter contributions
to the bare NGFP action and those resulting from integrating them out will in general no longer be complete.
On the EAA side, this situation is described by a trajectory $k\mapsto\Gamma_k$ that runs away from the fixed point
as $k$ decreases, and typically the resulting ordinary effective action of the gravity+matter system, $\Gamma_{k=0}$,
will indeed be affected by the presence of matter.

This expected behavior seems to be matched by the results of very recent 2D Monte-Carlo simulations of CDT coupled to
more than one massive scalar field \cite{AGJZ14}. It was found that, above a certain value of their mass, the dynamics
of the CDT+matter system is significantly different from the massless case. In particular, a characteristic
``blob + stalk'' behavior was observed, well known from 4D pure gravity CDT simulations, but absent in 2D with
conformal matter.

\section{Conclusions}
\label{sec:Conc}

In this paper we started from the Einstein--Hilbert truncation for the effective average action of metric quantum
gravity in $d>2$ dimensions and constructed its intrinsically $2$-dimensional limit. Contrary to earlier work on the
$\ve$-expansion of $\beta$-functions this limit was taken directly at the level of the action functional. We saw that
it turns the (local, second-derivative) Einstein--Hilbert term into the non-local Polyakov action.

Using this result we were able to conclude that in 2D the non-Gaussian fixed point underlying Asymptotic Safety gives
rise to a \emph{unitary} conformal field theory whose gravitational sector possesses the central charge $+25$. We
analyzed the properties of the fixed point CFT using both a gauge invariant description and a calculation based on the
conformal gauge where it is represented by a Liouville theory.

Finding that the complete fixed point action amounts to a CFT with vanishing total central charge, we compared and
contrasted asymptotically safe quantum gravity in $2$ dimensions with non-critical string theory.

Furthermore, exploring the gravitational dressing in 2D asymptotically safe gravity coupled to conformal matter we
discovered a rather surprising compensation mechanism that leads to a complete quenching of the KPZ-scaling,
that is, of the behavior we would have expected to occur a priori. Remarkably enough, this quenching is precisely what
is observed in Monte-Carlo simulations of analogous systems in the framework of causal dynamical triangulation. We
argued that these observations can possibly be interpreted as a reflection of the compensating effect displayed by
the integrated-out matter fluctuations and the explicit dependence of the bare gravitational action on properties of
the matter sector.

We close with a number of further comments.

\medskip
\noindent
\textbf{(1)}
An important step in proving the viability of the Asymptotic Safety program consists in demonstrating that Hilbert
space positivity can be achieved together with Background Independence and the nonperturbative renormalizability. While
we consider our present result on the unitarity of the pertinent CFT as an encouraging first insight, it is clear,
however, that the 2D case is not yet a crucial test since the gravitational field has no independent degrees of
freedom, and so there is no pure-gravity subspace of physical states whose positivity would be at stake. To tackle
the higher dimensional case additional techniques will have to be developed. Nevertheless, it is interesting that at
least at the purely geometric level the remarkable link between the Einstein--Hilbert and the Polyakov action which we
exploited has an analogue in all even dimensions $d=2n$. Each nontrivial cocycle of the Weyl cohomology yields, in an
appropriate limit $d\rightarrow 2n$, a well-defined non-local action that is known to be part of the standard effective
action in $2n$ dimensions \cite{MM01}.

\medskip
\noindent
\textbf{(2)}
A number of general lessons we learned here will be relevant in higher dimensions, too. We mention in particular that
the issue of unitarity cannot be settled by superficially checking for the stability of some bare action and ruling out
``wrong sign'' kinetic terms as this is implied sometimes. We saw that the CFT which is at the heart of the NGFP is
unitary \emph{even though} in conformal gauge it entails a negative kinetic energy of the Liouville field.
As we explained in section \ref{sec:NGFPCFT}, the background field, indispensable in our approach to quantum gravity,
plays an important role in reconciling these properties.

\medskip
\noindent
\textbf{(3)}
We determined the crucial central charge $\cgr$ from the leading term in the $\beta$-function of Newton's constant, and
we saw that depending on which parametrization of the metric is chosen the pure gravity result is either $25$ or $19$
for the exponential and the linear parametrization, respectively. In section \ref{sec:Emergence} we found convincing
evidence for accepting the result of the former, $+25$, as the correct one in the present context. Nevertheless, the
issue of parametrization dependence is not fully settled yet, and one should still be open towards the possibility that
the two sets of results, obtained from the same truncation ansatz but different choices of the fluctuating field,
actually might refer to \emph{different universality classes}.

\medskip
\noindent
\textbf{(4)}
Regarding different universality classes, it is perhaps not a pure coincidence that
the ``$19$'' is also among the ``critical dimensions for non-critical strings'' which were found by Gervais
\cite{G90}:
\begin{equation}
 D_\text{crit}=7,13,19.
\end{equation}
They correspond to gravitational central charges $c_\text{grav}=19,13,7$,
respectively. For these special values the Virasoro algebra admits a unitary truncation, that is, there exists a
subspace of the usual state space on which a corresponding chiral algebra closes, and which is positive (in the sense
that it contains no vectors $|\psi\rangle$ with $\langle\psi|\psi\rangle<0$). The associated string theories were
advocated as consistent extensions of standard Liouville theory, which is valid only for $c<1$ and $c>25$ when gravity
is weakly coupled, into the strongly coupled regime, $1<c<25$, in which the KPZ formulae would lead to meaningless
complex answers.

Thus, for the time being, we cannot exclude the possibility that a better understanding of the RG flow computed with
the linear parametrization (but with more general truncations than those analyzed in the present paper) will lead to
the picture that there exists a second pure gravity fixed point compatible with Hilbert space positivity, namely at
$c_\text{grav}=19$, and that this fixed point represents another, inequivalent universality class.

We know already that this picture displays the following correlation between pa\-ra\-me\-tri\-za\-tion and universality
class, which we would then indeed consider the natural one: The exponential parametrization, i.e.\ the ``conservative''
one in the sense that it covers only nonzero, nondegenerate, hence ``more classical'' metrics having a fixed signature,
leads to $c_\text{grav}=25$ which is located just at the boundary of the strong coupling interval. In the way it is
employed, the linear parametrization, instead, gives rise to an integration also over degenerate, even vanishing tensor
field configurations not corresponding to any classical metric; typically enough, it
is this parametrization that would be linked to the hypothetical, certainly quite non-classical theory with
$c_\text{grav}=19$ deep in the strong coupling domain.

Whatever the final answer will be it seems premature, also in more than $2$ dimensions, to regard the exponential
parametrization merely as a tool to do calculations in a more precise or more convenient way than this would be
possible with the linear one. It might rather be that in this manner we are actually computing \emph{something else}.

\vspace{4em}
\noindent
\textbf{Acknowledgments}:
We are grateful to J.~Ambj\o{}rn, J.~Gizbert-Studnicki, R.~Loll, C.~Pagani and R.~Percacci for helpful discussions.

\clearpage
\appendix

\section{Weyl transformations, zero modes and the induced gravity action}
\label{app:Weyl}

In this appendix we list the behavior of various geometric objects under Weyl transformations, including the induced
gravity functional, which is needed in the main part of this paper. Weyl transformations are given by
\begin{equation}
 g_\mn = \e^{2\sigma} \hg_\mn \,,
\label{eq:DefWeylTransf}
\end{equation}
where $\sigma$ is a scalar function on the spacetime manifold.

\medskip
\noindent
\textbf{(1)}
From the definition of the Christoffel connection we immediately obtain
\begin{equation}
 \Gamma^\alpha_\mn=\hat{\Gamma}^\alpha_\mn + \delta^\alpha_\mu \hD_\nu\sigma + \delta^\alpha_\nu \hD_\mu\sigma
  - \hg_\mn\hD^\alpha\sigma\, .
\label{eq:WeylChrist}
\end{equation}
Note that indices (on the right hand side) are raised and lowered by means of $\hg^\mn$ and $\hg_\mn$, respectively.
From \eqref{eq:WeylChrist} we easily deduce the Riemann tensor and its contractions,
\begin{align}
\begin{split} R^\alpha_{\mu\nu\rho} &= \hR^\alpha_{\mu\nu\rho} + 2\,\hg_{\mu[\nu}\hD_{\rho]}\hD^\alpha \sigma
  - 2\,\delta^\alpha_{[\nu}\hD_{\rho]}\hD_\mu \sigma - 2\,\hg_{\mu[\nu}\hD_{\rho]}\sigma \hD^\alpha \sigma\\
  &\hspace{3.34em}+ 2\,\delta^\alpha_{[\nu}\hD_{\rho]}\sigma \hD_\mu \sigma
  + 2\,\hg_{\mu[\nu}\delta^\alpha_{\rho]} \hD_\beta \sigma \hD^\beta \sigma \, ,
\end{split}\\
R_\mn &=  \hR_\mn - (d-2)\Big(\hD_\mu \hD_\nu \sigma - \hD_\mu \sigma \hD_\nu \sigma\Big)
   - \hg_\mn \Big[\hB\sigma +(d-2)\hD_\alpha\sigma\hD^\alpha \sigma \Big] , \\
R &= \e^{-2\sigma}\left[\hR-(d-1)(d-2)\hD_\mu\sigma \hD^\mu\sigma -2(d-1)\hB\sigma\right] ,
\end{align}
where $\hB\equiv\hD_\alpha\hD^\alpha$ and the square brackets enclosing indices denote antisymmetrization,
$A_{[\mn]} = \frac{1}{2}(A_\mn-A_{\nu\mu})$. Note that since the underlying connection is the Christoffel symbol,
i.e.\ it is torsion free, we have $\hD_\mu\hD_\nu \sigma = \hD_\nu\hD_\mu \sigma$. For the Einstein tensor we have
\begin{equation}
 G_\mn = \hat{G}_\mn + (d-2)\left[ -\hD_\mu \hD_\nu \sigma + \hg_\mn\hB\sigma + \hD_\mu\sigma \hD_\nu\sigma
  + \frac{d-3}{2}\hg_\mn \hD_\alpha\sigma \hD^\alpha\sigma \right].
\label{eq:WeylEinstein}
\end{equation}
Furthermore, the metric determinant transforms as
\begin{equation}
 \sg = \shg\, \e^{d\sigma} \, .
\end{equation}
Hence, we arrive at the useful relations
\begin{align}
 \sg\,R &= \e^{(d-2)\sigma}\shg \left[\hR-(d-1)(d-2)\hD_\mu\sigma \hD^\mu\sigma-2(d-1)\hB\sigma\right], \\
 \int\dd x\sg\,R &= \int\dd x\shg\,\e^{(d-2)\sigma} \left[\hR+(d-1)(d-2)\hD_\mu\sigma \hD^\mu\sigma
 \right].\label{eq:EHexpanded}
\end{align}
The transformation behavior of the Laplacian is given by
\begin{equation}
\Box f = \e^{-2\sigma}\hB f +(d-2)\e^{-2\sigma} \hD_\mu \sigma \hD^\mu \sigma \, ,
\label{eq:WeylLaplace}
\end{equation}
where $f$ is an arbitrary scalar function.

\medskip
\noindent
\textbf{(2)}
In the special case of two dimensions, $d=2$, we obtain
\begin{align}
 R &= \e^{-2\sigma}\left[\hR-2\hB\sigma\right],
 \label{eq:WeylR}\\
 \sg\,R &= \shg \left[\hR-2\hB\sigma\right],
 \label{eq:WeylgR}\\
 \Box f &= \e^{-2\sigma}\hB f\,.
\end{align}

\medskip
\noindent
\textbf{(3)}
Due to its relevance for the induced gravity action we are particularly interested
in the transformation behavior of $\Box^{-1}R\mku$, with the inverse Laplacian (Green function)
$\Box^{-1}\equiv\Box^{-1}(x,y)$, where $(\Box^{-1}R)(x)$ refers to
\begin{equation}
 (\Box^{-1}R)(x)=\int\dd y\sg\;\Box^{-1}(x,y)R(y).
\label{eq:BoxMOne}
\end{equation}
If $\Box$ has no zero modes, its inverse
is defined by $\Box\big[\Box^{-1}(x,y)\big]=\frac{1}{\sg}\delta(x-y)$. On the other hand, if $\Box$ has normalizable
zero modes, then $\Box^{-1}$ is defined as the inverse of $\Box$ on the orthogonal complement to its kernel, where the
delta function has to be modified appropriately, that is, $\Box\,\Box^{-1}(x,y)=\frac{1}{\sg}\delta(x-y)-
\text{Pr}_0(x,y)$ where $\text{Pr}_0$ denotes the projection onto zero modes. Whenever we write $\Box^{-1}$ in this
article, this definition is meant implicitly.

\medskip
\noindent
\textbf{(4)}
Since the consideration of zero modes requires a more careful treatment, we first consider the situation where zero
modes are absent in the following subsection, before investigating the general case in subsection \ref{app:Zero}.

\subsection{The induced gravity action in the absence of zero modes}
\label{app:NoZero}

If the Laplacian has no zero modes, then the equation $\Box f=h$ can be solved for $f$ by direct inversion of $\Box$,
that is, $f=\Box^{-1}h$. In this case the transformation behavior of the Green function $\Box^{-1}$ is given by
\begin{equation}
 \Box^{-1}\big(\e^{-2\sigma}\,h\big) = \hB^{-1}h\,.
\end{equation}
This gives rise to
\begin{equation}
 \Box^{-1}R = \hB^{-1} \hR -2\sigma.
 \label{eq:WeylInvBoxR}
\end{equation}

For our arguments in section \ref{sec:EpsilonLimit} we need to determine the transformation behavior of the induced
gravity functional $I[g]$ which can be defined as the normalized finite part of Polyakov's induced effective action
\cite{P81}
\begin{equation}
\textstyle
 \Gi[g] = \frac{1}{2}\Tr\ln(-\Box)\,.
\label{eq:GiWithoutZero}
\end{equation}
In the absence of zero modes the trace in \eqref{eq:GiWithoutZero} can be computed explicitly. The result, $\Gi[g]$,
consists of a universal finite part and a regularization scheme dependent divergent part. Regularizing by means of a
proper time cutoff \cite{BV87}, for instance, one obtains from eq.\ \eqref{eq:GiWithoutZero}:
\begin{equation}
 \Gi[g] = \frac{1}{96\pi}\int\td^2 x\sg\, R\,\Box^{-1}R - \frac{1}{8\pi s}\int\td^2 x\sg\,.
\label{eq:GammaInd}
\end{equation}
The second term on the RHS of eq.\ \eqref{eq:GammaInd} is scheme dependent and divergent in the limit
$s\rightarrow 0$. It might be absorbed by a redefinition of the cosmological constant. The first term, on the other
hand, contains all relevant information, so we focus on it for our further investigations. We define the induced
gravity functional $I[g]$ to be proportional to \emph{the finite part of} $\Gi[g]$,
\begin{equation}
 I[g] \equiv 96\pi\; \Gi[g]\big|_\text{finite} = \int\td^2 x\sg\, R\,\Box^{-1}R \,.
\label{eq:IDefinition}
\end{equation}
Using \eqref{eq:WeylgR} and \eqref{eq:WeylInvBoxR} we now obtain, after integrating by parts,
\begin{equation}
  I[g] = \int\td^2x\shg\left[\hR\,\hB^{-1}\hR-4\hR\sigma+4\sigma\hB\sigma\right].
\end{equation}
This can be written as
\begin{equation}[b]
 I[g] - I[\hg] = - 8\,\Delta I[\sigma;\hg],
\label{eq:ItoDeltaI}
\end{equation}
with the functional $\Delta I$ defined by
\begin{equation}
 \Delta I[\sigma;\hg] \equiv \frac{1}{2}\int\td^2x\shg\left[\hD_\mu\sigma\hD^\mu\sigma+\hR\sigma\right].
\label{eq:DeltaI}
\end{equation}

These results prove useful for calculating the 2D limit of the Einstein--Hilbert action, as applied
in sections \ref{sec:genRem} and \ref{sec:EpsilonLimit}.

\subsection{The treatment of zero modes}
\label{app:Zero}
What is different and which results of section \ref{app:NoZero} remain valid when the scalar Laplacian has one or more
zero modes? To illustrate the issue let us start from scratch and consider a functional integral over a simple scalar
field $X$ minimally coupled to the metric. Integrating out $X$ will ``induce'' a gravity action for the metric then.
The corresponding partition function is given by
\begin{equation}
 \tilde{Z}[g] \equiv \int\mD X\;\e^{-\frac{1}{2}\int\td^2 x\sg\,g^\mn\mku\p_\mu X\,\p_\nu X}
 = \int\mD X\;\e^{-\frac{1}{2}\int\td^2 x\sg\,X(-\Box) X}\,.
\label{eq:ZDiv}
\end{equation}
(The notation with the tilde is chosen since definition \eqref{eq:ZDiv} is pathological and gets modified as shown
in the following.) Let us expand the field $X$ in terms of normalized eigenmodes $\vp^{(n)}$ of the Laplacian
$-\Box$, that is, $X = \sum_n c_n\, \vp^{(n)}$, where $-\Box\mku \vp^{(n)}=\lambda_n\vp^{(n)}$, with the normalization
$\int\td^2 x\sg\;\vp^{(n)}(x)\,\vp^{(m)}(x)=\delta_{mn}$. Then the integral in \eqref{eq:ZDiv} can be written as
\begin{equation}
 \tilde{Z}[g] = \int\prod\limits_n \frac{\td c_n}{\sqrt{2\pi}}\;\e^{-\frac{1}{2}\sum_n \lambda_n\,c_n^2}\,.
\end{equation}
Now let us suppose that the Laplacian has a zero mode, $-\Box\mku\vp^{(0)}=0$, i.e.\ $\lambda_0=0$. In this case the
integration over its Fourier coefficient, $\int\td c_0 \;\e^{-\frac{1}{2}\lambda_0\,c_0^2} =\int\td c_0\;1$, is
\emph{divergent}, and so is $\tilde{Z}[g]$. Thus, the zero mode(s) has to be \emph{excluded from the path integral}
in the first place. The correct definition reads
\begin{equation}
 Z[g] \equiv \int\mD' X\;\e^{-\frac{1}{2}\int\td^2 x\sg\,g^\mn\mku\p_\mu X\,\p_\nu X}\,.
\label{eq:ZCorrect}
\end{equation}
Here and in the following the prime denotes the exclusion of zero modes.

We will consider only connected manifolds with vanishing boundary. In that case the Laplacian has (at most) one single
normalized zero mode. It is given by
\begin{equation}
 \vp^{(0)} = 1/\sqrt{V}\,,
\end{equation}
with the volume, or area, $V=\int\td^2 x\sg\,$.

Performing the Gaussian integrals in eq.\ \eqref{eq:ZCorrect} one obtains
\begin{equation}
 Z[g]=\big[\detp(-\Box)\big]^{-\frac{1}{2}}.
\label{eq:ZTildeComp}
\end{equation}
The corresponding effective action $\Gi$ is determined by $Z\equiv\e^{-\Gi}$, leading to
\begin{equation}
\textstyle
 \Gi[g] = \frac{1}{2}\ln\detp(-\Box) = \frac{1}{2}\Trp\ln(-\Box)\,,
\label{eq:GiDef}
\end{equation}
which is Polyakov's induced gravity action, adapted to taking account of zero modes. In order to find an integral
representation for $\Gi$ similar to eq.\ \eqref{eq:GammaInd} it turns out convenient to consider the variation of $\Gi$
under a finite Weyl transformation, giving rise to a strictly local term and a term involving the logarithm of the
volume (see e.g.\ \cite{FV11}): The finite part of the variation reads
\begin{equation}
 \Gi[g] - \Gi[\hg] = - \frac{1}{12\pi}\,\Delta I[\sigma;\hg] + \frac{1}{2}\,\ln\left(V/\hat{V}\right) ,
\label{eq:GitoDeltaIZero}
\end{equation}
with the volume terms $V\equiv\int\td^2 x\sg$ and $\hat{V}\equiv\int\td^2 x\shg$, and with $\Delta I[\sigma;\hg]$ as
defined in eq.\ \eqref{eq:DeltaI}. The second term on the RHS of \eqref{eq:GitoDeltaIZero} originates from the zero
mode contribution contained in the conformal factor.

To extract from \eqref{eq:GitoDeltaIZero} an explicit expression for $\Gi$ that depends only on one metric, we would
like to eliminate the conformal factor and rewrite also the RHS of \eqref{eq:GitoDeltaIZero} as the difference between
some functional evaluated at $g$ and the same functional evaluated at $\hg$.
Although the existence of such a representation can be proven \cite{Do94},
the explicit form of $\Gi[g]$ with only one argument is (to the best of our knowledge) not known in general.
As already pointed out in ref.\ \cite{Du94}, the problem occurs for uniform rescalings when the conformal factor is a
constant, i.e.\ proportional to the zero mode: In this case even the formula $\int g^\mn\,\frac{\delta S[g]}{\delta
g_\mn}= \frac{1}{2}\frac{\p S[\e^{2\sigma}g]}{\p \sigma}\big|_{\sigma=0}\,$, where $\sigma$ is a constant, does not
apply, a counterexample being the induced gravity functional \eqref{eq:IDefinition} which is invariant under uniform
rescalings but whose metric variation gives rise to the anomaly proportional to $R$.

To eliminate the conformal factor in \eqref{eq:GitoDeltaIZero} we would like to solve the equation
\begin{equation}
 \Box\mku\sigma = \textstyle\frac{1}{2\,\sg}\left(\shg\hR-\sg R\right)
\label{eq:sigmaInTermsOfgandhg}
\end{equation}
for $\sigma$, where \eqref{eq:sigmaInTermsOfgandhg} follows from \eqref{eq:WeylgR} and the identity
$\shg\,\hB = \sg\,\Box$, valid in 2D. The existence of a solution is guaranteed by the fact that the RHS of
\eqref{eq:sigmaInTermsOfgandhg} is orthogonal to the zero mode, thanks to topological invariance. 
However, the conformal factor itself could have a contribution from the zero mode. As a consequence, the solution
for $\sigma$ is not unique. Employing the Green function $\Box^{-1}$ as defined below eq.\ \eqref{eq:BoxMOne}
we obtain
\begin{equation}
 \sigma = \textstyle\frac{1}{2}\,\Box^{-1}\frac{1}{\sg}\left(\shg\hR-\sg R\right) + \frac{1}{V}\int\sg\,\sigma,
\label{eq:SolSigma}
\end{equation}
where the second term is the constant zero mode part. (Recall that $\Box^{-1}$ is the inverse of $\Box$ on the
orthogonal complement to the kernel of $\Box$, and it satisfies $\Box\,\Box^{-1}(x,y)=\frac{1}{\sg}\delta(x-y)-
\frac{1}{V}$.)
Making use of the relation $\sigma = \frac{1}{2}\ln(\sg/\shg)$ the last term in \eqref{eq:SolSigma} can be expressed
in terms of the metrics $g_\mn$ and $\hg_\mn$, too. Then eq.\ \eqref{eq:GitoDeltaIZero} becomes
\begin{equation}
 \Gi[g]-\Gi[\hg] = \Gi[g,\hg],
\end{equation}
with the both $g_\mn$- and $\hg_\mn$-dependent functional \cite{Do94}
\begin{equation}
\begin{split}
 \Gi[g,\hg] \equiv\; &\frac{1}{96\pi}\int\textstyle\left(\sg R+\shg\hR\right)\Box^{-1}\frac{1}{\sg}
 \left(\sg R-\shg\hR\right)\\
 &-\textstyle\frac{\chi}{12\mku V}\int\sg\,\ln\left(\frac{\sg}{\shg}\right)
 +\frac{1}{2}\,\ln\left(\frac{V}{\hat{V}}\right).
\end{split}
\label{eq:IWeylTransGeneral}
\end{equation}
In this expression it does not seem possible to disentangle $g$ from $\hg$.

Nevertheless, by introducing a fiducial metric $g_0$ in \eqref{eq:IWeylTransGeneral} we could define $\Gi[g]$ formally
up to an additive constant by
\begin{equation}
 \Gi[g]\equiv \Gi[g,g_0].
\end{equation}
Using this definition, $\Gi$ satisfies indeed eq.\ \eqref{eq:GitoDeltaIZero}. The corresponding functional
$I_\text{full}[g]$ (where $I_\text{full}$ refers to the general case, with zero mode and arbitrary rescalings) can be
obtained by applying rule \eqref{eq:IDefinition}, $I_\text{full}[g] \equiv 96\pi\; \Gi[g]|_\text{finite}\,$,
resulting in
\begin{equation}
 I_\text{full}[g] \equiv I[g] + R[g,g_0],
\label{eq:DefIFull}
\end{equation}
with $I[g]=\int\!\sg\,R\,\Box^{-1}R$ as above, and with the residue
\begin{equation}
 R[g,g_0] \equiv -\int\sqrt{g_0}\,R(g_0)\Box^{-1}{\textstyle\frac{\sqrt{g_0}}{\sg}}\,R(g_0)
 - \frac{8\pi\chi}{V}\int\sg\,\ln\left(\textstyle\frac{\sg}{\sqrt{g_0}}\right)
 +48\pi\,\ln\left(\textstyle\frac{V}{V_0}\right).
\label{eq:Residue}
\end{equation}
This residue is due to the zero mode contribution to the conformal factor relating $g$ with $g_0$.
Using eq.\ \eqref{eq:GitoDeltaIZero} leads to a transformation behavior of $I_\text{full}[g]$ similar to the one found
in section \ref{app:NoZero}. We obtain
\begin{equation}[b]
 I_\text{full}[g] - I_\text{full}[\hg] = - 8\,\Delta I[\sigma;\hg] + 48\pi\,\ln\left(V/\hat{V}\right) .
\label{eq:ItoDeltaIZero}
\end{equation}
Thus, apart from the pure volume terms we recover the same result as in eq.\ \eqref{eq:ItoDeltaI}, the modification
being due to the zero modes of $\Box$ and $\hB$, $\vp^{(0)}=1/\sqrt{V}$ and $\hat{\vp}^{(0)}=1/\sqrt{\hat{V}}$,
respectively.

Concerning our results of section \ref{sec:IndGravityFromEH} we observe that $I[g]$ is to be replaced according to
\begin{equation}
 I[g] \;\rightarrow\; I_\text{full}[g]-48\pi\,\ln(V/V_0),
\label{eq:IReplace}
\end{equation}
where the corresponding behavior under Weyl transformations is given by eq.\ \eqref{eq:ItoDeltaIZero}. Thus, in the
general case there are additional correction terms in consequence of the zero modes. In particular,
eq.\ \eqref{eq:LimitResult} generalizes to
\begin{equation}
 \frac{1}{\ve}\int\td^{2+\ve}x\sg\, R
 = -\frac{1}{4}I[g] + Q[g,g_0] +\frac{4\pi\chi}{\ve}+C\big(\{\tau\}\big)+\mO(\ve),
\label{eq:LimitResultGen}
\end{equation}
with the correction terms $Q[g,g_0]\equiv\frac{1}{4}\int\!\sqrt{g_0}\,R(g_0)\Box^{-1}\frac{\sqrt{g_0}}{\sg}\,R(g_0)
 + \frac{2\pi\chi}{V}\int\!\sg\,\ln\left(\frac{\sg}{\sqrt{g_0}}\right)$. We would like to point out that the crucial
result in eq.\ \eqref{eq:LimitResult}, the appearance of the non-local action $I[g]$, is contained in its extension
\eqref{eq:LimitResultGen}, too. All conclusions in the main part of this paper that relied on the emergence of
$I[g]$ in the 2D limit of the Einstein--Hilbert action remain valid in the presence of zero modes. The correction terms
in \eqref{eq:LimitResultGen} do not change our main results, in particular the central charge, which is read off from
the prefactor of $I[g]$, remains unaltered.

Finally, two comments are in order.

\medskip
\noindent
\textbf{(1) Nonvanishing Euler characteristics}. We would like to point out the following subtlety concerning the
induced gravity functional $I$. As argued above, $\Box^{-1}$ is defined such that it affects only nonzero modes while
it ``projects away'' the zero modes of the objects it acts on. In particular, the function $(\Box^{-1}R)(x)$ satisfies
$\Box\,\Box^{-1}R = R - \frac{1}{V}\,4\pi\chi$. Hence, for manifolds with vanishing Euler characteristic, $\chi=0$,
we recover the usual feature of an inverse operator, $\Box\,\Box^{-1}R=R$, as long as $\Box^{-1}$ acts on $R$. The
reason behind this property is that the Fourier expansion of $R$ cannot contain any contribution $\propto c_0\vp^{(0)}$
from the zero mode if $\chi=0$. As a consequence $\Box^{-1}R$ is non-zero provided that $R$ does not vanish, and, in
turn, $I[g]$ is a non-zero functional.

On the other
hand, if $\chi\neq 0$, then it might happen that $I[g]$ vanishes. As an example, let us consider a sphere with constant
curvature $R>0$. Since $R$ is proportional to the constant zero mode in this case, we have $\Box^{-1}R=0$, and thus
$I[g]=0$. With regard to eq.\ \eqref{eq:ItoDeltaIZero} this means that all nontrivial contributions to the LHS must
come from $I_\text{full}[\hg]$ and from the residue contained in $I_\text{full}[g]$.

\medskip
\noindent
\textbf{(2) A modified induced gravity functional}.
The occurrence of the volume term in eq.\ \eqref{eq:ItoDeltaIZero} can be understood as follows. We removed the zero
modes from the path integral \eqref{eq:ZCorrect}, and this exclusion affects the transformation behavior,
replacing \eqref{eq:ItoDeltaI} with \eqref{eq:ItoDeltaIZero}. However, there is the possibility to redefine the
partition function in order to absorb the volume terms. Let us briefly sketch the idea.

As above we expand the scalar field $X$ in the partition function in terms of normalized eigenmodes $\vp^{(n)}$ of
the Laplacian, $X=\sum_n c_n\,\vp^{(n)}$,
and insert this into eq.\ \eqref{eq:ZCorrect}. Then it is easy to show (see e.g.\ \cite{ID89}) that the transformation
behavior of $\ln Z$ under a Weyl variation according to eq.\ \eqref{eq:DefWeylTransf}, $\delta g_\mn=2\sigma\, g_\mn$,
is given by
\begin{equation}
 \delta\ln Z =\int\td^2 x\sg\,\left(\frac{1}{4}\frac{\delta g}{g}\right)\sum_{n=0}^\infty\big[\vp^{(n)}\big]^2
 -\frac{1}{2}\frac{\delta V}{V}\,.
\end{equation}
Rearranging terms yields
\begin{equation}
 \delta\ln\left(\sqrt{V/V_0}\,Z\right)
 = \int\td^2 x\sg\,\left(\frac{1}{4}\frac{\delta g}{g}\right)\sum_{n=0}^\infty\big[\vp^{(n)}\big]^2\,,
\label{eq:Rearranged}
\end{equation}
where $V_0$ is an arbitrary reference volume introduced merely to render the argument of the logarithm dimensionless.
The advantage of eq.\ \eqref{eq:Rearranged} is that its RHS does no longer contain any distinction between zero and
nonzero modes, so the combination $\sqrt{V/V_0}\,Z$ is more appropriate for a treatment of all modes on an equal
footing.

These observations suggest introducing the modified definition
\begin{equation}
 Z_\text{mod}[g] \equiv \sqrt{V/V_0}\int\mD' X\;\e^{-\frac{1}{2}\int\td^2 x\sg\,g^\mn\mku\p_\mu X\,\p_\nu X}\,.
\label{eq:ZDef}
\end{equation}
The corresponding effective action reads
\begin{equation}
\textstyle
 \Gamma_\text{ind,mod}[g] = \frac{1}{2}\ln\detp(-\Box)-\frac{1}{2}\ln\frac{V}{V_0} \,.
\label{eq:GiMod}
\end{equation}
This modified effective action is often used in the literature \cite{CKHT87}. Applying the rule \eqref{eq:IDefinition}
to \eqref{eq:GiMod} and using \eqref{eq:DefIFull} yields the \emph{modified induced gravity functional}
\begin{equation}
 I_\text{mod}[g] \equiv I_\text{full}[g]-48\pi\,\ln{\textstyle\frac{V}{V_0}} \,,
\end{equation}
consistent with \eqref{eq:IReplace}.
Employing eq.\ \eqref{eq:ItoDeltaIZero} we find that it transforms according to
\begin{equation}[b]
 I_\text{mod}[g] - I_\text{mod}[\hg] = - 8\,\Delta I[\sigma;\hg],
\end{equation}
with $\Delta I$ as defined in eq.\ \eqref{eq:DeltaI}. Thus, for $I_\text{mod}[g]$ we recover the same behavior under
Weyl transformations as for $I[g]$ in eq.\ \eqref{eq:ItoDeltaI}, which was the transformation law for the case without
zero modes.

In conclusion, zero modes can be taken into account by employing a modified definition of the path integral, where the
behavior of the (generalized) induced gravity functional under Weyl rescalings remains essentially the same.

\end{spacing}


\end{document}